\documentstyle[11pt]{article}

\def\afour{
\setlength{\topmargin}{0mm}
\setlength{\headheight}{0mm}
\setlength{\headsep}{0mm}
\setlength{\textwidth}{6in}
\setlength{\textheight}{248mm}
\setlength{\oddsidemargin}{.25in}
\setlength{\evensidemargin}{.25in}
}

\newcommand\eq[1]{Eq.~(\ref{#1})}
\newcommand\eqs[2]{Eqs.~(\ref{#1}) and (\ref{#2})}
\newcommand\eqss[3]{Eqs.~(\ref{#1}), (\ref{#2}) and (\ref{#3})}

\newcommand\eqst[2]{Eqs.~(\ref{#1})--(\ref{#2})}
 
\newcommand\rfrac[2]{\left(\frac{#1}{#2}\right)}
 
\newcommand{\sub}[1]{_{\mbox{\scriptsize#1}}}
\newcommand{\su}[1]{^{\mbox{\scriptsize#1}}}

\newcommand\ee{\end{equation}}
\newcommand\be{\begin{equation}}
\newcommand\eea{\end{eqnarray}}
\newcommand\bea{\begin{eqnarray}}
 
\newcommand\sunit{\,\mbox{sec}}

\newcommand\km{\,\mbox{km}}

\newcommand\GeV{\,\mbox{GeV}}

\newcommand\Mpc{\,\mbox{Mpc}}
 
\newcommand\mone{^{-1}}
\newcommand\mtwo{^{-2}}
\newcommand\mthree{^{-3}}

\newcommand\mhalf{^{-1/2}}

\newcommand\half{^{1/2}}

\newcommand\quarter{^{1/4}}
 
\newcommand\msun{M_\odot}
\newcommand\mpl{m_{Pl}}
 
\newcommand\pa{\partial}
\newcommand\pdif[2]{\frac{\pa #1}{\pa #2}}
 
\newcommand\lsim{\mathrel{\rlap{\lower4pt\hbox{\hskip1pt$\sim$}}
    \raise1pt\hbox{$<$}}}
\newcommand\gsim{\mathrel{\rlap{\lower4pt\hbox{\hskip1pt$\sim$}}
    \raise1pt\hbox{$>$}}}
 
\newcommand\diff{\mbox d}

\def\calp{{\cal P}}
\def\calr{{\cal R}}
 
\afour
 
\newcommand\del{{\mbox{\boldmath $\nabla$}}}

\newcommand\bfk{\mbox{\bf k}}

\newcommand\bfr{\mbox{\bf r}}
\newcommand\bfv{\mbox{\bf v}}
\newcommand\bfw{\mbox{\bf w}}
\newcommand\bfe{\mbox{\bf e}}
\newcommand\bfx{\mbox{\bf x}}

\newcommand\bfu{\mbox{\bf u}}
\newcommand\sk{_{\mbox{\scriptsize \bf k}}}

\def\calr{{\cal R}}

\afour
 
\begin{document}
 
\begin{flushright}
LANCS-TH 94-15\\
SUSSEX-AST 94/8-2\\
astro-ph/9408066\\
(August 1994)\\
\end{flushright}
\begin{center}
\Large
{\bf Interpreting Large Scale Structure Observations}
 
\vspace{.3in}
\normalsize
\large{David H. Lyth$^{\dagger}$ and Andrew R. Liddle$^*$} \\
\normalsize
 
\vspace{.6 cm}
{\em $^{\dagger}$School of Physics and Materials, \\ University of
Lancaster, \\ Lancaster LA1 4YB.~~~U.~K.}\\
 
\vspace{.4cm}
{\em  $^*$Astronomy Centre, \\ Division of Physics and Astronomy, \\
University of Sussex, \\ Brighton BN1 9QH.~~~U.~K.}\\
 
\vspace{.6cm}
To appear in the proceedings of\\
``Journee Cosmologie'', held in Paris, June 1994.\\
Presented by D.~H.~Lyth
 
\vspace{.6 cm}
{\bf Abstract}
\end{center}
 
\vspace{.2cm}
\noindent
 
The standard model of large scale structure is considered, in which
the structure originates as a Gaussian adiabatic density perturbation
with a nearly scale invariant spectrum. The basic theoretical tool
of cosmological perturbation theory is described, as well as 
the possible origin of the density perturbation as a vacuum fluctuation 
during inflation. Then, after normalising the spectrum to fit the cosmic 
microwave background anisotropy measured by COBE, some versions of the 
standard
model are compared with a variety of data coming from observations of 
galaxies and galaxy clusters. The recent COBE analysis of G\'{o}rski
and collaborators is used, which gives a significantly higher normalization
than earlier ones. The comparison with galaxy and cluster data is done using 
linear theory, supplemented by the Press-Schechter formula when discussing 
object abundances of rich clusters and of damped Lyman alpha systems. By
focussing on the smoothed density contrast as a function of scale, the 
observational data can be conveniently illustrated on a single 
figure, facilitating easy comparison with theory. The spectral index is 
constrained to $0.6<n<1.1$, and in particle physics motivated models that 
predict significant gravitational waves the lower limit is tightened to 
$0.8$.
 
\section{Introduction}
 
By `large scale structure' one means galaxies and clusters, with emphasis on 
their spatial distribution and motions, and also the cosmic microwave 
background (cmb) anisotropy. We are at present 
on the verge of a quantum leap in our understanding of large scale structure, 
because the cmb anisotropy is being measured for the first time
\cite{smet,gorski,scottrev,scottwhite}. In particular, we may in the 
forseeable future verify or rule out
the standard model of structure formation, according to which large 
scale structure arises from a Gaussian, adiabatic density perturbation
that is nearly scale invariant at horizon entry. If that model is 
verified there will be  the dazzling prospect of a window on
the fundamental interactions on scales approaching the Planck scale,
because the vacuum fluctuation during inflation generates just 
such a perturbation.
 
This article has two objectives.
One is to equip the non-specialist with a starting point, from which to 
follow the saga that will unfold during the coming years.
To this end we include an extensive discussion of linear 
perturbation theory, which allows one to translate theoretical input coming 
from say a model of inflation into a form amenable for comparison with 
observations. Our other aim
is to provide a critical assessment of the present position,
in the light of all relevant observations, including the abundance of 
damped Lyman alpha systems and the latest analysis of the crucial
COBE data. These and the other relevant observations are summarized on a 
single plot. On the same plot are drawn the canonical version of the 
standard model, and variants which alter the
Hubble constant $h$, the spectral index $n$, and the fraction 
$\Omega_\nu$ of any hot dark matter. We discuss the observational
constraints on these three parameters, and 
in particular the constraint on $n$, which is of great interest
because it can be a sensitive discriminator between inflationary models.
As we have emphasised several times in earlier publications
\cite{will,LL1,LL2,mymnras,berkeley,capri} a powerful constraint on
$n$ is provided by the 
long `lever arm' between the very large scale explored by the 
large angle cmb anisotropy and the smaller scales explored by
galaxy and cluster data. On the basis of the presently available data
we find that $0.6<n<1.1$, with lower limit tightened to $0.8$ in
particle physics motivated models of inflation that 
generate significant gravitational waves. The lower limit in particular
is rather firm because several different types 
of observation confirm it.
 
\subsection*{Three possible models for large scale structure}
 
Three possible models of large scale structure are commonly entertained. 
The standard model is that it originates as a Gaussian adiabatic density 
perturbation, whose spectrum is nearly scale independent at horizon entry.
This model has been explored far more 
thoroughly than the other two, and is the only one that will be 
considered here. An alternative is that large scale structure originates from
topological defects, such as cosmic strings.
In both of these models the underlying scale invariance means that
galaxy and cluster formation is directly related to the 
magnitude of the large scale cmb anisotropy. It also means that
the first objects to form are the progenitors of galaxies.
Finally there is the possibility that large scale structure 
originates from a density perturbation (either adiabatic or 
isocurvature) with a spectrum that is not even approximately scale 
independent. In that case structure formation could be very different,
with perhaps much lighter objects forming first, and it would not be 
directly related to the magnitude of 
the large scale cmb anisotropy.
 
The standard model has two features which distinguish it from the others, and 
make it so attractive. One of them concerns theory. If, as is widely 
supposed, the initial conditions for the hot big bang are set by inflation, 
then an adiabatic, Gaussian, more or less scale invariant density 
perturbation is {\em predicted}. What could be more natural than to suppose 
that perturbation will explain 
large scale structure? If that turns out to be so, large scale 
structure will provide us with a unique window on the nature of the
fundamental interactions, because both the magnitude and precise scale 
dependence of the perturbation are highly model dependent
\cite{LL1,paul,davesal,mymnras}.
 
The other feature 
concerns phenomenology. The model has been intensively studied 
and is relatively simple, so that by now one knows 
how to estimate its predictions for most available types of data.
To a first approximation the predictions depend on
a single number, specifying the magnitude of the density perturbation
at horizon entry. If that number is chosen to fit 
the cmb anisotropy, all other data can certainly be explained to within a 
factor of two or three! Of course the burning question 
is whether the data can actually be explained within their observational 
uncertainties, which in the best cases are only tens of percent.
The answer to that question depends what other parameters are available in 
the standard model, and it will be our main focus.
 
\subsection*{An overview of the standard model}
 
The standard model assumes that large scale 
structure originates as an adiabatic, Gaussian density perturbation 
whose spectrum is more or less scale-independent at horizon entry.
The scale dependence (if any) is parameterised as a power law, with a 
spectral
index that by convention is defined so that $n=1$ corresponds to 
scale invariance. 
 
In order to have any chance of agreeing with observation, the standard 
model requires non-baryonic dark matter, which is
more or less 
cold, and which has a density dominating the baryon density.
By `cold' one means that the constituent particles are 
stable, non-interacting and non-relativistic, at all relevant epochs.
 
The simplest version of the standard model assumes that $n=1$,
and that the
(non-baryonic) 
dark matter is completely cold. It also assumes that the
energy density of the universe is critical, $\Omega=1$,
with no cosmological constant or other exotic contribution
so that there is the standard cosmology with matter domination
at present.
This critical density, $n=1$ CDM model
was the favoured one for 
many years. It contains only two free parameters, which are the 
normalisation of the spectrum, and the value of the 
Hubble constant $H_0\equiv100h\km\sunit\mone \Mpc\mone$.
According to
observations having nothing to do with large scale structure (namely, 
direct observations and measurements of the age of the universe),
$0.4\lsim h\lsim 0.6$.\footnote{Higher 
values of $h$ are permitted only in low density models, especially those 
featuring a cosmological constant. Given $h$, the baryon density is 
practically fixed by the standard nucleosynthesis relation $\Omega_B 
h^2=0.013\pm 0.002$.} 
Assuming the central value $h=0.5$ we arrive at 
a canonical version of the CDM model,
 whose only free parameter is the magnitude of the scale 
invariant density perturbation.
 
The canonical CDM model does not agree with observation 
because the predicted spectrum of the
density perturbation has the wrong scale 
dependence. On large scales the spectrum is accurately determined by
the COBE measurement of the cmb anisotropy, and
as one goes down in scale through the regime explored by data on galaxies 
and clusters it becomes progressively too big compared with 
the data. How can we reduce the small scale power?
 
The simplest possibility is to reduce $h$ below the 
canonical value $h=0.5$, which reduces the small scale density 
perturbation 
because it delays matter domination giving the density perturbation
less time to grow. It has recently been noted \cite{lowh}
that this `old universe' option might work if $h$ is as low as $0.3$,
but such a value is
difficult to reconcile with measurements of $h$
from Hubble's law.
 
Another possibility is to reduce $n$ below the canonical value $n=1$. 
This `tilted spectrum' option has been widely investigated  
\cite{will,LL1,tilt,natural2,gelb,LL2,berkeley,capri},
and it might be viable with $n\simeq 0.7$. 
Unfortunately most inflation models with tilt also generate 
significant gravitational waves which ruin this concordance, as we discuss 
later. Of course one
can combine the old universe and tilted spectrum options.
 
A third option is to change the hypothesis of completely
cold dark matter, which reduces the small scale 
density perturbation 
because cold dark matter maximises its rate of growth.
An attractive possibility, from both a theoretical and 
observational viewpoint, is to invoke a fraction $\Omega_\nu$ of hot dark 
matter in the form of massive neutrinos. 
This {\it mixed dark matter} (MDM) model has 
been widely investigated
\cite{mdm,lymanalpha,davis,pogosyan,bobqaisar,silvio,mymnras,boblast,capri}, 
both with and without the option of allowing $h$ and $n$
to depart from their canonical values. We shall see that 
it may be observationally viable with $\Omega_\nu=0.15$ or so 
and the canonical $h$ and $n$.
Other possibilities yet to be investigated fully
are to replace the CDM by some form of warm dark matter
like sterile neutrinos, or to replace it by
decaying or self-interacting dark matter.
 
The final possibility is to reduce the matter density,
either  with \cite{lambda,both,tegsilkopen}
or without \cite{lowomega,both,tegsilkopen} a cosmological 
constant to keep the total energy density critical.
In practice one takes the non-baryonic dark matter to be completely cold
in that case. This
{\it low density CDM model}
has been quite widely investigated and 
with $h$ and $n$ at their canonical values
it may be observationally viable with $\Omega_c$
of order $0.5$ or so.
 
All of this assumes that there is no significant gravitational wave 
contribution to the cmb anisotropy. A contribution up to 50\% or so is not 
ruled out by present data, and is actually predicted by some models of 
inflation \cite{LL1,paul,davesal}.
Of the models that are well motivated from particle 
physics, those predicting a significant contribution also predict a
spectral index $n<1$, and on the large scales explored by COBE they predict 
that the relative gravitational wave contribution to the mean square
cmb anisotropy is $R\simeq 6(1-n)$. In these models the normalization of the 
density perturbation is therefore reduced by a factor $[1+6(1-n)]\mhalf$, and 
the possible existence of this factor should be taken into account when 
considering tilt in the spectral index. On the other hand, almost 
all inflation models suggest that if the spectral index is very close to 1 
then the gravitational wave contribution will be negligible.
 
In this article the focus is on the (critical density)
MDM model, which of course includes the critical density
CDM model as a special case, allowing
$h$ and $n$ to vary, and keeping in mind the 
possibility of gravitational waves.
The low density CDM model will be mentioned only briefly,
and other possibilities not at all.
 
\section{Cosmological perturbations}
 
In this section we give a theoretical overview of cosmological 
perturbation theory, which will later be amplified 
and compared with observation. We also derive briefly 
the inflationary predictions for large scale structure, which 
will provide a unique 
window on the fundamental interactions if the standard 
model is verified.
 
\subsection{Linear cosmological perturbation theory}
 
The foundation for the standard model of structure formation is linear
cosmological perturbation theory, according to which the perturbations 
satisfy a set of linear partial differential equations as long as 
they are sufficiently small
 \cite{lifs,peebles80,synch,bardeen,kosa,hawk,ellis,lyth85,brly,lymu,lyst}. 
Using linear theory, one can follow the growth of an enhancement 
in the mass density until it begins to collapse to form a gravitationally 
bound structure. One can also give an almost complete description of the cmb 
anisotropy.
 
The perturbations decouple into three separate modes,
traditionally called scalar, vector and tensor modes. Vector 
perturbations are associated with the vorticity of the fluid flow,
and are presumably negligible since the vorticity decays. Tensor
perturbations are, for practical purposes, freely propagating
gravitational waves, and so are very easy to handle. 
They are predicted at some level by inflation, and might 
affect the cmb anisotropy.
 
This leaves the scalar perturbations. They are completely defined by giving, 
for each particle species, the perturbation in the momentum distribution 
function\footnote
{In principle the 
perturbations in the momentum distribution functions have vector and tensor 
modes as well as the scalar one being considered here, but the vector
mode is supposed to be absent, and the tensor mode is negligible because 
gravitational waves have negligible coupling to matter.}. 
From this
one can calculate the perturbation in the matter density, pressure and 
anisotropic stress of each species. (It turns out these in turn determine the 
metric perturbation, but the latter is not directly observable.)
 
For a particle species whose collisions are sufficiently frequent to 
make it a perfect fluid, the full momentum distribution 
is not needed and one needs only the energy density and pressure.
This is a useful approximation for baryons, electrons and 
photons before photon decoupling. Alternatively, if a particle species 
has negligible random motion it is automatically a perfect fluid with
negligible pressure, and one then needs only its mass density. 
This is the case for cold dark matter and for baryons well after 
decoupling. These perfect fluid approximations provide roughly correct 
results for both the matter and the radiation, but are not adequate
to address the increasingly accurate observations now becoming 
available.
 
In order to define a momentum distribution function one needs
a set of worldlines, specifying the observers who measure the momentum.
Also, to define the perturbation in this or any other quantity
one needs to slice spacetime into hypersurfaces so that the quantity
can be 
split up into an average plus a perturbation.\footnote
{The only exception is if the unperturbed quantity (ie., the average)
is time independent, in which case the perturbation is 
independent of the slicing to first order.}
Any choice of worldlines and slicing
will do, but a convenient one that we adopt here is
to use comoving observers and the hypersurfaces orthogonal to them, 
which are called comoving hypersurfaces.
(Comoving observers by definition move with the total energy flow,
which means that they measure zero total momentum density.)
 
\subsubsection*{Independent Fourier modes}
 
It is convenient to expand each perturbation as a Fourier series
in comoving coordinates, because in linear theory each comoving Fourier mode
evolves independently of the others. 
Comoving Cartesian coordinates $\bfx$ are related to physical 
Cartesian coordinates
$\bfr$ by $\bfr = a\bfx$, where $a(t)$ is the scale factor of 
the universe. It will be convenient to normalise $a=1$ at present.
A generic perturbation will be denoted by $f$,
and the Fourier series is defined
in a comoving box much bigger than the observable universe,
\be
f(\bfx)=\sum\sk f\sk \exp(i\bfk.\bfx)
\label{116}
\;.\ee
For each Fourier mode the
physical wavenumber is $k/a$ (where $k=|\bfk|$),
 and its present value is just $k$.
At any epoch, a feature in a perturbation $f$ with size
$a R$ is described by modes 
with wavenumber of order $\sim a/k$.
 
\subsubsection*{Before horizon entry}
 
A crucial epoch for each scale is horizon entry, 
when the inverse wavenumber $a/k$ first falls 
within the Hubble distance $H\mone$. 
For critical density, the  scale entering the horizon
at matter-radiation equality is $k\mone=(20h\mone)h\mone\Mpc$.
Smaller scales enter the horizon during radiation domination.
 
Before horizon entry there is no time for causal processes to operate, and 
each comoving region of the universe evolves like a separate Friedmann 
universe \cite{lyth85,lymu}.
 The evolution of the perturbations is therefore very simple,
and can be calculated just by comparing the independent evolution of the 
separate Friedmann universes. 
 
From now on, we specialise to the standard model of structure formation.
The standard model states that the initial perturbations 
are adiabatic and hence determined by the curvature perturbation,
and it also states that the curvature perturbation is Gaussian and
has a more or less scale independent spectrum. 
 
\subsubsection*{The adiabatic initial condition}
 
The adiabatic initial condition
states that well before horizon entry
the momentum distribution function of each species is a 
function only of the total energy density, or equivalently of the 
temperature.
For a perfect fluid species it is enough to specify the 
energy density and pressure, and for a non-relativistic
(matter) species the mass density is enough. What functions of
total energy density these quantities are
depends on the assumed cosmology, the usual assumption being 
that radiation (extreme relativistic) species are in thermal equilibrium 
and have practically zero chemical potential.
 
The adiabatic initial condition determines all the initial
perturbations in terms of the 
energy density perturbation. The perturbations are indeed
adiabatic, because 
the evolution of every quantity 
along each comoving worldline is that of a Friedmann 
universe which is adiabatic. (This evolution defines the perturbations, 
because they all vanish on a hypersurface of constant energy density.)
From the  relations $\rho_m\propto a^3$
and $\rho_r\propto a^4$ for radiation and matter in a Friedmann
universe, one learns that the 
density contrasts of the radiation and matter 
species are related by
\be
\frac{\delta\rho_m}{\rho_m}=\frac34
\frac{\delta\rho_r}{\rho_r}
\label{2}
\;.\ee
 
In a Friedmann universe  the curvature 
(measured in a comoving distance unit)
is constant. One can therefore define a curvature
perturbation, which is constant on each scale well before horizon entry
\cite{lyth85}.
It is conveniently specified by the quantity
\be
\calr=-H\delta t
\label{rdef}
\ee
where
$\delta t$ is the displacement of the comoving hypersurfaces from
flat hypersurfaces \cite{LL2}.
On each scale $\calr\sk$ is constant well before 
horizon entry, and it 
determines the density perturbation $\delta\rho\sk$.
At horizon entry $\delta\rho\sk/\rho\simeq\calr\sk$,
so one can think of the curvature perturbation as specifying the 
magnitude of the density contrast at horizon entry.
 
\subsubsection*{Gaussian perturbations}
 
In comparing theory with observation, one is interested only in the 
stochastic properties of the perturbations. According to the standard
model the initial curvature perturbation is
Gaussian, which means essentially that its Fourier 
components have random phases (apart from the 
reality condition $f\sk=f^*_{\mbox{-\scriptsize \bfk}})$.
By virtue of the adiabatic condition,
this Gaussian property is bequeathed to all of the perturbations.
 
All stochastic properties of a Gaussian perturbation are 
determined by its spectrum, defined as 
the modulus squared of its Fourier component. To be precise,
we define the
spectrum $\calp_f$ of a generic perturbation $f$ as
\be
\calp_f(k)
=4\pi(Lk/2\pi)^3\langle |f\sk|^2\rangle
\label{4}
\ee
where $L$ is the comoving size of the box used for the Fourier expansion,
and $\langle\rangle$ indicates that $|f\sk|^2$ has been averaged over a small 
region of $k$ space to make it smooth. The spectrum is independent of 
the direction of $\bfk$ because no direction in space is preferred.
The normalization is chosen so that 
according to Parseval's theorem the mean square perturbation $\sigma_f^2$ is 
\be
\sigma_f^2=\int^\infty_0 \calp_f(k) \diff k/k
\;.\ee
Evaluated at random positions, each perturbation has a Gaussian 
probability distribution with variance $\sigma_f^2$.
The spectrum $\calp_T(l)$ of the cmb anisotropy is defined in an analogous
way, as essentially the  modulus squared of its $l$th multipole.
 
\subsubsection*{Scale independence of the initial spectrum}
 
The linear evolution gives all of the spectra in terms of the spectrum 
$\calp_\calr$ of the initial curvature perturbation. According to the 
standard model $\calp_\calr$
is more or less scale dependent, any scale dependence being 
parameterised by a power law $\calp_\calr\propto k^{n-1}$. (The appearance of 
$n-1$ instead of $n$ is a historical accident.) Exact scale independence 
corresponds to a spectral index $n=1$, whereas $n<1$ corresponds to a 
`tilted' spectrum which has less power on small scales. One can also 
contemplate a `blue' spectrum \cite{LL1,latest,sylvia},
 $n>1$, but this possibility is not favoured by 
either theory or observation.
 
\subsubsection*{The transfer functions}
 
Except on very large scales, the perturbations evolve in a complicated 
manner after horizon entry, as the particles move about under the combined 
effect of gravity and particle collisions. By the present epoch however,
the situation has again become simple. On cosmologically interesting 
scales the scale dependence of the matter density perturbation becomes 
fixed, and so does that of the anisotropy of the cmb.
The spectrum of the density contrast is related to $\calp_\calr$
by a linear transfer function, and so is the spectrum of the
cmb anisotropy. The transfer functions depend in general on
the value of $h$, on the nature and the amount of the 
non-baryonic dark matter and on 
the value of the cosmological constant
if one is introduced. The transfer function for the cmb anisotropy 
also depends on the epoch of re-ionisation if it is late.
 
\subsubsection*{Observable scales}
 
Observations of large scale structure probe a range
of scales. 
 
The scales probed by the cmb anisotropy are easy to estimate.
It originates at the surface of last scattering, whose distance 
is now close to $2H_0\mone=6000h\mone\Mpc$, 
and the thickness of the last scattering surface is
about $7h\mone\Mpc$. This means that in round figure it probes scales
$10\Mpc \lsim k\mone \lsim 10^4\Mpc$.
 
Working out the scales that are probed by galaxy and cluster 
observations is more difficult.
An upper limit of order $100\Mpc$ comes just from the fact that 
much more 
distant regions of the universe have yet to be surveyed in detail,
but to describe the lower limit we have to describe structure formation.
 
In viable versions of the standard model, copious
structure formation begins only at a redshift of a few, 
when objects with a very broad range of masses form
at more or less the same epoch. In order for an object to be reasonably 
stable,
the baryons within it have to be able to collapse and
dissipate their energy, which is thought to be possible only
for masses bigger than $10^6\msun$ or so. The resulting objects
are thought to become early galaxies.
Afterwards there is a bottom-up picture of structure formation,
successively more massive objects forming in 
sequence ending at the present epoch with
rich galaxy clusters of mass around $10^{15}\msun$.
The lower limit of the mass range to which the bottom-up picture applies 
is not very well defined but something like $10^{11}\msun$
is a reasonable order of magnitude to have in mind \cite{LL2}.
 
In the bottom-up picture, each object originates as a slightly overdense 
region in the early universe, the region attracting more matter 
until it eventually collapses under its own weight. 
As long as its overdensity is small, the region expands with the universe
and is described by linear theory. The Fourier
modes describing it 
have inverse wavenumber $a/k$ of order its size
and in a critical density universe the mass contained in a sphere with 
radius $a/k$ is
\be
M(k\mone) =1.16\times 10^{12} h^2 (k\mone/1\Mpc)^3 \msun
\label{76}
\;.\ee
Thus, observations relating to the bottom-up picture of structure formation
explore scales in the range
$1\Mpc\lsim k\mone\lsim 10\Mpc$.
 
What are the relevant observations? Since
big clusters formed recently (in the standard model),
each of them can probably be identified 
with one of 
the first objects that formed in this mass range. The same may be true
of smaller clusters and `groups', with masses $10^{13}\msun\lsim M
\lsim 10^{14}\msun$. What about smaller masses? Present day galaxies have 
$10^6\msun\lsim M\lsim 10^{12}\msun$, but because
of merging and other astrophysics they 
can probably not be identified with the first objects forming in this 
mass range. However one also observes 
quasars and damped Lyman alpha systems
out to a redshift of 3 or 4. 
Their masses are thought to be in roughly the range 
$\msun\sim 10^{11}\msun$ to $10^{13}\msun$,
and they can probably be identified with  the 
first objects forming in this mass range.
The formation of the first objects with mass below this range
is not directly observed,
but observational constraints on the epoch of re-ionisation 
should provide an indirect probe in the forseeable future.
 
The conclusion is that one can observe the formation of the first 
structure with mass from $10^{15}\msun$ down to perhaps
$10^{11}\msun$,
thereby exploring the density perturbation on scales $1\Mpc\lsim k\mone\lsim 
10\Mpc$. Bigger scales than this can be explored by looking at the spatial
distribution and motion of galaxies and clusters.
Smaller scales cannot be explored through present observations
because one does not understand galaxy evolution,
but in the near future the cmb anisotropy 
may constrain the epoch of re-ionisation and provide an indirect probe 
on the scale $.01\Mpc$.
 
\subsection{The inflationary prediction for the initial spectrum}
 
The original motivation \cite{harrison} for scale independence
was simply that it is rather natural, particularly in view of the fact 
that $\calp_\calr(k)$ is essentially the spectrum of the density 
contrast at the epoch of horizon entry, when non-trivial evolution 
begins. If inflation sets the initial conditions, approximate
scale dependence is however {\em predicted}
\cite{adpred,lyth85,mukhanov,sasaki,stly}. Let us see briefly how this 
works.
 
During inflation the energy density is supposed to be dominated by the scalar 
field potential $V(\phi)$, and $\Omega$ is driven to 1 
so that\footnote
{Einstein gravity is assumed, which can usually be achieved 
by a suitable definition of the relevant fields.}
\be 
H^2\simeq \frac13 \frac{8\pi}{\mpl^2} V  \label{288}
\ee
where $\mpl^2=G\mone$ is the Planck mass and as usual 
we are setting $c=\hbar=1$.
The spectral index depends only
on the first and second derivatives of $V$ which
are conveniently defined by
the dimensionless parameters
\bea
\epsilon &\equiv& \frac{m_{Pl}^2}{16\pi}
\left( \frac{V'}{V} \right)^2 \label{287}\\
\eta &\equiv& \frac{m_{Pl}^2}{8\pi} \frac{V''}{V} \label{291}
\eea
where primes denote derivatives with respect to $\phi$. Inflation 
typically occurs only if 
\be
\epsilon\ll 1,\hspace{10mm}|\eta| \ll 1
\label{flat}
\;.\ee
This implies that $H$ is slowly 
varying (on the Hubble timescale) so that 
\be
a\propto \exp(Ht)
\label{exp}
\;.\ee
Independently of its initial
value, $\dot\phi$ typically settles down to the
`slow-roll' approximation
\be
3H\dot\phi=-V'
\label{slowroll}
\;.\ee
 
On any spatial hypersurface the 
gradient of $\delta\phi$ gives the momentum 
density, so $\delta\phi$ vanishes on the 
comoving hypersurfaces. The
curvature of these hypersurfaces is defined 
in terms of their displacement $\delta t(\bfx)$
from flat hypersurfaces by \eq{rdef}.
On flat hypersurfaces the inflaton field perturbation is therefore
$\delta\phi=-\dot\phi\delta t=\dot\phi\calr/H$. 
 
In typical models of inflation $\delta\phi$ has negligible interaction 
with itself and other fields, so that each Fourier mode evolves 
independently (consistent with the assumption that
linear cosmological perturbation theory applies during inflation).
Quantization is straightforward and the mass of $\delta \phi$ is 
typically negligible. In order for inflation to be useful 
cosmologically interesting scales have to start well within the 
horizon,
and the corresponding Fourier modes of every field must be in the 
vacuum or the corresponding particles would
dominate the energy density and spoil inflation \cite{LL2}.\footnote
{We are here making the usual assumption that
the curvature is negligible on the present Hubble scale, ie., 
that the density is critical counting the contribution of any
cosmological constant. The assumption 
can be relaxed, but then one cannot invoke flat spacetime field theory to 
define the vacuum. There is however a natural definition,
and if it is used and the inflaton field is assumed to be in the vacuum 
the formulas below still hold \cite{lystomega}.}
The vacuum expectation value of $|\delta \phi\sk|^2$ 
is therefore 
given by the usual flat spacetime calculation (the scalar field
version of 
the Casimir effect that is experimentally verified for the 
electromagnetic field). It can be evolved forward in time using the 
classical equation of motion that $\delta\phi\sk$ satisfies 
in the Heisenberg representation, and using \eqs{flat}{slowroll}
one obtains the result
\cite{mukhanov,lyth85,sasaki,stly}
\be
\calp_\phi\half\simeq H/2\pi
\label{phifluc}
\;.\ee
 Since $H$ and $\dot\phi$ are slowly varying
it follows that the curvature spectrum a few Hubble times 
after horizon exit is given by
\be
\calp_\calr\half(k) \simeq \frac{H^2_*}{2\pi\dot\phi_*}
\label{13}
\ee
where the star denotes the epoch of horizon exit $aH=k$.
This value is retained until the scale $k$ enters the horizon 
again, after inflation has ended. Using \eq{288}
it can be written
\be
\calp_\calr(k)\simeq \frac1{24\pi^2}\rfrac{8\pi}{\mpl^2}^2
\frac{V_*}{\epsilon_*}
\;.\ee
The observed normalization $\calp_\calr\half\simeq 5.8\times10^{-5}$
is a powerful constraint on models of inflation. In particular,
since $\epsilon_*\ll1$ the energy scale $V_*\quarter$ must be 
less than about $10^{16}\GeV$ and so must the temperature after 
inflation. This last requirement makes life quite difficult for cosmic 
string theories of large scale structure though the situation is not
hopeless \cite{latest}.
 
Using \eqs{flat}{slowroll} the
spectral index $n=\diff \log\calp_\calr/\diff \log k$ is given by
\cite{LL1,stly} 
\be
n(k)-1\simeq 2\eta_*-6\epsilon_*
\;.\ee
In typical models of inflation
$\eta$ and $\epsilon$ vary slowly on the Hubble
timescale. Since cosmologically interesting scales 
range over only about four decades of $k$, this means that 
$n(k)$ can be regarded as a 
constant leading to the
power law
$\calp_\calr\propto k^{n-1}$.
 
Contrary to the view that is sometimes expressed, there is nothing
very controversial about this calculation. 
Provided that $\delta\phi$ has negligible interactions
its vacuum fluctuation well before horizon entry is 
a standard result of flat spacetime field theory, and its subsequent 
evolution is given by the classical equation of motion. If 
\eqs{flat}{slowroll}
are valid the quoted result is obtained,
but if these conditions fail one can still calculate the 
evolution in a straightforward manner using for instance the 
formalism described in \cite{stly}. The predicted spectrum will 
then generally have a strong departure from scale invariance and the model 
will be in danger of being ruled out by observation.
Even the assumption of negligible interaction can 
perhaps also be relaxed,
leading to a non-Gaussian perturbation 
\cite{nongauss}.
Again, the non-Gaussianity endangers the model.
 
A similar calculation \cite{starob,LL2,stly} for the gravitational waves 
gives their relative contribution to the spectrum of the 
cmb anisotropy, which on large scales
is $R\simeq 12\epsilon$. For the particle physics motivated models in which 
$R$ is significant, $V(\phi)$ is well approximated by an exponential or 
high power-law, which gives $\eta\simeq2\epsilon$ and hence
$R\simeq6(1-n)$.
 
Measured values of $1-n$ or $R$ significantly different from zero
would provide a unique window on the inflationary potential, and hence 
on the fundamental interactions. Even the upper bounds on these 
quantities that already exist rule out the usual versions of 
`extended inflation', because \cite{LL1}
such models give both $n\lsim 0.7$ and
$R\simeq 6(1-n)$.
 
Now we study in turn the three observable perturbations, which are the 
matter density perturbation, the peculiar velocity and the cmb anisotropy.
 
\section{The Matter Density Perturbation}
 
In the Newtonian regime cosmological perturbation theory is
a straightforward application of fluid flow equations,
taking into account where necessary particle diffusion and 
free-streaming (collisionless particle movement). 
On large scales and in the early universe the Newtonian treatment is 
inadequate and must be replaced by one based on general relativity,
but here too it is possible to derive fluid flow equations
using a treatment strongly resembling the Newtonian one
\cite{hawk,ellis,lyth85,lymu,lyst,brly}. We employ that treatment here
instead of the more usual metric perturbation approach
\cite{lifs,bardeen,kosa}.\footnote
{The same equations can be derived using the `gauge invariant'
version of the metric perturbation approach \cite{bardeen,kosa},
but then the physical significance of the quantities is obscured, and
one loses the connection with the Newtonian treatment \cite{brly}.}
 
\subsection{The fluid flow equations}
 
The strategy is to populate the universe with comoving observers, who define 
physical quantities in their own region. 
A crucial concept is the {\it velocity gradient} $u_{ij}$. It is
defined by each comoving observer, using locally inertial coordinates in 
which the observer is instantaneously at rest, as the gradient of
the velocity $u^i$ of nearby comoving observer,
\be
u_{ij}\equiv \partial_j u_i
\label{70}
\;.\ee
(Note that $u_i$ is defined locally, with only its gradient having global 
significance.) The velocity gradient can be uniquely decomposed into an 
antisymmetric vorticity $\omega_{ij}$, a symmetric traceless shear 
$\sigma_{ij}$, and a locally defined Hubble parameter $H$,
\be
u_{ij}=H \delta_{ij}+\sigma_{ij} +\omega_{ij}\;. \label{71}
\ee
 
In the limit of homogeneity and isotropy, $\sigma_{ij}=\omega_{ij}=0$.
Then the universe is homogeneous on comoving hypersurfaces (those 
orthogonal to comoving worldlines).\footnote
{As noted in Section 3 this definition has to 
be modified if the vorticity is not negligible.}
 The continuity equation is
\be
\frac{\diff \rho}{\diff \tau}
=-3H(\rho+p)
\label{contin}\;.
\ee
where $\rho$ is the energy density, 
$p$ is the pressure and $\tau$ is the proper time measured
by comoving observers. The deceleration of gravity is given by
\be
\frac{\diff H}{\diff \tau}=
-H^2-\frac{4\pi G}{3} (\rho+3p) 
\;.\ee
These lead to the Friedmann equation
\be
H^2=\frac{8\pi G}{3} \rho-\frac{K}{a^2}
\label{fried}
\;.\ee
The constant $K$ is a measure of the curvature of comoving
hypersurfaces,
 and we assume critical density $\Omega=1$ 
which corresponds to $K=0$.
 
Now consider perturbations.
One can show from angular momentum conservation that $\omega_{ij}$
decays like $(\rho+p)a^{-5}$.
It is therefore  presumably negligible
though we shall see in the next section how it may be handled if 
desired. We will also see there how to calculate $\sigma_{ij}$
which is {\em not} negligible, but for the moment we 
need only $H$.
 
On a given comoving hypersurface, each
quantity can be split into
an average plus a perturbation,
\bea \rho(\bfx,t)&=& \bar\rho(t)+\delta\rho(\bfx,t) 
\label{110} \\
p(\bfx,t)&=& \bar p(t)+\delta p(\bfx,t) 
\label{110a} \\
H(\bfx,t)&=& \bar H(t)+\delta H(\bfx,t) 
\label{110b} \;.\eea
The time coordinate $t$ labels 
the hypersurfaces, and will be taken to be the average of $\tau$.
We would like to choose the space coordinates
${\bfx}=(x^1,x^2,x^3)$ to be
comoving coordinates, related to Cartesian coordinates by $r^i=a x^i$, with
$a$ the average scale factor given by $\dot a/a=\bar 
H$. This cannot be done exactly, because
the expansion is not isotropic and the
comoving hypersurfaces are not flat.  However these effects are
of first order, and can therefore be ignored
when describing perturbations which are themselves of first order. In other
words {\it all perturbations `live' in the unperturbed Friedmann 
universe.} 
 
Along each comoving worldline the continuity equation \eq{contin}
is unchanged. Ignoring aniso\-tropic stress the
Friedmann equation receives extra terms to become
\cite{lymu,lyst},
\be
\frac{\diff H}{\diff \tau}=
-H^2-\frac{4\pi G}{3} (\rho+3p) -\frac13 \frac
	{\nabla^2 \delta p}{\rho +p}
+\mbox{(anisotropic stress term)}
 \label{raych} 
\;.\ee
This equation is called the Raychaudhuri equation. 
The anisotropic stress term is typically of the same order of magnitude 
as the exhibited one, or smaller.
 
\subsection{Separate Friedmann universes}
 
If the additional terms 
in \eq{raych} are negligible, the evolution along each 
comoving worldline is identical with that in a Friedmann universe. The 
Friedmann equation \eq{fried} is obeyed, with some $K$ that is 
constant along each comoving worldline. Assuming critical density its average 
vanishes, so we write $K\equiv \delta K(\bfx)$ to remind ourselves that it is 
a perturbation.
 
To see when the additional terms will be 
negligible it is enough to compare the exhibited term with
the density perturbation $-4\pi 
G\delta\rho/3$ coming from the second term
(it is straightforward to 
verify that the latter does not cancel with other terms,
and we have just noted that the anisotropic stress term is
not expected to dominate the exhibited one).
For a given Fourier component $\nabla^2=-(k/a)^2$, 
and since the density is critical the relative contribution of the
exhibited term is $\sim (k/aH)^2 (\delta p\sk/\delta\rho\sk)$.
Well before horizon entry the adiabatic condition ensures that 
$\delta p\sk\lsim\delta \rho\sk$, and
the relative contribution is small. It is also small sufficiently long 
after matter domination, because then $\delta p$ becomes negligible.
The conclusion is that {\em on each scale 
a comoving region evolves like a Friedman 
universe before horizon entry, and begins to do so again
sufficiently long after horizon entry.}
 
Given this local Friedmann evolution one can calculate
the density perturbation in terms of the curvature perturbation
$\calr$ \cite{lyth85}. It can be shown \cite{bardeen,LL2} 
that the definition \eq{rdef} is equivalent 
to\footnote
{The factor $k^2$ means that
${\cal 
R} \sk = (3/2)
(\delta K\sk/a^2) (a^2/k^2)$ measures the curvature perturbation 
in units
of the relevant scale $a/k$. 
(The intrinsic curvature scalar of the comoving hypersurfaces is
$R^{(3)}=6K/a^2$.)
Yet another interpretation of $\calr$ is that it 
is essentially the Newtonian gravitational potential caused by $\delta
\rho$.}
\be {{\cal R}} \sk \equiv\frac32 \frac{\delta K\sk }{k^2}
\label{90}\;.\ee
Perturbing the Friedmann equation gives, to first order,
\be 2H\delta H\sk =\frac{8\pi G}{3} \delta \rho \sk 
-\frac {\delta K\sk }{a^2} \label{25}  \;.\ee
We also need to perturb the continuity equation \eq{contin}.
In doing so 
one has to remember that the comoving worldlines are not geodesics
because of the pressure gradient. As a result,
the proper time interval $d\tau$ between a pair of
comoving hypersurfaces is position dependent. 
One can show (using essentially the Lorentz transformation
between nearby observers) that
its variation with position is given by \cite{lyst,LL2}
\be \frac{d\tau}{dt}=\left(1-\frac{\delta p}{\rho+p} \right)
\label{111} \;.\ee
Taking this into account, perturbing the continuity equation gives
\be
\frac{\diff \delta \rho\sk}{\diff t}=-3(\rho+p)\delta H\sk-3H \delta\rho\sk 
\label{27} 
\;.\ee
Eliminating $\delta H\sk$ from \eqs{25}{27} then gives
for the density contrast $\delta=\delta\rho/\rho$
\be \frac{2H\mone}{5+3w} \frac{d}{dt}
\left[ \rfrac{aH}{k}^2 \delta\sk \right]
+ \rfrac{aH}{k}^2 \delta\sk 
=\frac{2+2w}{5+3w} {{\cal R}}\sk  \label{121} \ee
where $w=p/\rho$.
During any era when $w$ is constant, \eq{121} has the solution
(dropping a decaying mode)
\be \rfrac{aH}{k}^2 \delta\sk =\frac{2+2w}{5+3w} {{\cal R}} \sk
\label{123} \;.\ee
In the radiation dominated era before horizon entry this becomes
\be
\rfrac{aH}{k}^2 \delta\sk =\frac49 {{\cal R}}\sk (\mbox{initial})
\label{125} \ee
and in the matter dominated era it becomes
\be \rfrac{aH}{k}^2 \delta\sk = \frac25 {{\cal R}}\sk (\mbox{final})
\label{31} \;.\ee
As the labels imply, we are regarding the value of $\calr\sk$
during the first era as an `initial condition', which determines
its value during the `final' matter dominated era.
 
For future reference note that during matter domination,
$H\propto t\mone\propto a^{-3/2}$ and
\be
\delta\sk\propto a \hspace*{1cm}\mbox{(matter domination)}
\label{32}
\;.\ee
 
\subsubsection*{The Newtonian picture}
 
\eqs{fried}{raych}, which determine the evolution of the 
energy density perturbation 
are valid in the framework of general relativity.
For the matter dominated case $p\ll\rho$ they have however the same form 
as the Newtonian equations, and at a given epoch
the Newtonian picture indeed applies to the universe on scales that
are sufficiently far inside the horizon \cite{peebles80}.
Thus we have 
demonstrated that the Newtonian equations for the density perturbation
are valid during matter domination,
even on scales that are too large for the Newtonian picture itself to be 
valid. We shall shortly demonstrate that the same is true for the 
equations that describe the
peculiar velocity.

\subsection{The transfer function}
 
On scales $k\mone\gg(20h\mone)h\mone\Mpc$, horizon entry is long after 
matter domination so that the initial and final eras overlap
and ${{\cal R}}\sk (\mbox{initial})={{\cal R}}\sk (\mbox{final})$. 
On smaller scales there is a {\it transfer function} $T(k)$,
which may be  defined by 
\be {{\cal R}}\sk (\mbox{final})= T(k) {{\cal R}}\sk (\mbox{initial})
\label{127} \;.\ee
The corresponding density contrast is given by \eq{31},
and it applies to each type of matter \cite{lyst}. It is Gaussian, and
its spectrum is
\be
{{\cal P}}_\delta(k) =\frac4{25} \calp_\calr(k)
T^2(k) \rfrac{k}{aH}^4 
\label{23b} \;.\ee
 
The transfer function depends on the nature of the dark matter,
and on the value of the Hubble constant. The main physical effects are 
the following \cite{peebles80,synch,kosa}
\begin{itemize}
\item The cold dark matter density contrast grows, as overdense regions 
attract more matter towards them. The growth is slow before matter 
domination, but rapid thereafter.
\item The baryons, electrons and photons form a tightly coupled fluid
before decoupling at redshift 
$z\sim 1000$, whose density contrast oscillates as 
a standing wave. The oscillations are damped because the photons 
diffuse out of overdense regions, and they carry some of the baryons
with them. 
\item After decoupling the photons travel freely.
\item After decoupling, the baryon density contrast grows to match that 
of the cold dark matter on scales $\gsim 10^6\msun$, but continues to 
oscillate for a long time on smaller scales
(this may account for the absence
of galaxies with $M\lsim 10^6\msun$).
\item Massless neutrinos free stream (travel freely) out
of any density enhancement.
\item Any hot dark matter (massive neutrinos)  free streams
while it is relativistic, and then its density contrast grows.
\end{itemize}
 
On each scale, the transfer function is applicable after
the random particle motion 
has become negligible. In the absence of 
hot dark matter this happens
soon after decoupling (on scales
bigger than $10^6\msun$). In the presence of hot dark matter
it happens only after a redshift given by \cite{davis}
\be 
k\mone\gsim 0.11(1+z)\half\Omega_\nu \Mpc \label{202}\;.\ee
As we are interested only in scales $\gsim 1\Mpc$,
the random motion has become negligible by the present.
It is still marginally significant at $z\sim 4$ on the scales
relevant for the formation of the
quasars and damped Lyman alpha systems, but the effect is small compared 
with observational uncertainties and we shall not include it.
 
\subsection*{The smoothed density contrast}
 
The linear theory described so far applies only 
to small perturbations. Except perhaps at early times, this means that
one has to smooth all quantities on a suitably large scale before 
using it. Adopting the simple `top hat' prescription, 
the density contrast $\delta(\bfx)$ at each point is replaced by its average 
$\delta(\bfx,R)$ within a comoving sphere,
whose present radius $R$ defines the smoothing scale.
Instead of specifying $R$ 
one can specify the average mass $M(R)$ within the sphere,
which assuming critical matter density $\Omega_0=1$
is given by
\be
M(R) =1.16\times 10^{12} h^2 (R/1\Mpc)^3 \msun
\label{mass}
\;.\ee
Top hat smoothing multiplies each Fourier coefficient 
by the window function
\be
W(k R)=3\left(\frac{\sin(kR)}{(kR)^3}-\frac{\cos(kR)}{(kR)^2}
\right)
\;.\ee
It therefore filters out Fourier modes with $k\mone\lsim R$.
On the other hand, the spectrum
$\calp_\delta(k)$ of the unsmoothed density contrast
decreases rapidly as $k\mone$ increases
on scales $k\mone\gsim 1\Mpc$,
corresponding to the fact that the probability of finding  a feature 
with size $a/k$ decreases rapidly.
As a result,
most of the
structure in the smoothed density contrast at a given epoch has 
size of order $a R$,
for smoothing scales $R\gsim 1\Mpc$.
 
The mean square $\sigma^2(R)$ of the smoothed density contrast
is 
\be
\sigma^2(R)=\int^\infty_0 W^2(kR) \calp_\delta(k) \diff k/k
\label{38}
\;.\ee
At any epoch it can 
can be made arbitrarily small by choosing a sufficiently large
filtering scale $R$, but it grows with time.
As long as $\sigma(R)\lsim1$, linear theory applies except in those rare
regions of space where $\delta(\bfx,R)\gsim1$.
These regions correspond to gravitationally bound 
objects with mass of order $M(R)$ and linear theory does not apply 
to them, but since they are rare
they will not significantly affect the evolution of
$\sigma(R)$, or of the Fourier modes with $k\mone\gsim R$ that dominate 
it. When  $\sigma(R)$ grows to become of order 1 on the other hand, the 
formerly rare gravitationally bound regions become common, and
the evolution becomes non-linear
on scales $k\mone\sim R$. 
Since $\sigma(R)\propto a\propto (1+z)\mone$ (\eq{32}),
this occurs at a redshift given by
\be 1+z\sub{nl}(R)=\sigma_0(R) \label{274}\ee
where the $0$ denotes the present value of the {\em
linearly evolved } quantity.
As we shall see $z\sub{nl}$ is no more than a few
in observationally viable versions of the standard 
model, even on the smallest relevant scales.
Before this epoch gravitationally bound objects are rare,
and whether they are too rare to account for what is observed is one of the 
crucial issues that we shall address.
 
\section{The peculiar velocity}
 
Associated with the density perturbation is a peculiar velocity 
field, specifying the departure from uniform Hubble flow that occurs
as matter falls into overdense regions. 
 
\subsection{The Newtonian regime}
 
The Newtonian picture \cite{peebles80}
applies to the universe after matter domination
on scales sufficiently far inside the horizon, and 
in particular it applies at present to the sphere of radius
a few hundred Mpc around us within which detailed observations
of galaxies and clusters have been made .
 
In the Newtonian picture
there is a globally defined fluid velocity field $\bfu$, and 
choosing the reference frame so that $\bfu$ vanishes at the origin the
peculiar velocity field ${\mbox{\bf v}}$ is defined\footnote
{Up to a 
constant, which can be chosen so that the average of ${\mbox{\bf v}}$ 
vanishes.} by
\be 
{\mbox{\bf u}}({\mbox{\bf r}})-{\mbox{\bf u}}(0)= \bar H {\mbox{\bf r}}+ 
{\mbox{\bf v}}(
{\mbox{\bf r}})-{\mbox{\bf v}}(0)\;,
\label{203}
\;.\ee
An equivalent statement in terms of the velocity gradient \eq{71} is
\be
\delta u_{ij}=\partial_i v_j\;.
\label{9a}
\ee
where $\partial_i=a\mone\partial/\partial x^i$.
Like any vector field, $\bfv$ can be written
${\mbox{\bf v}}={\mbox{\bf v}}\su{L}+{\mbox{\bf v}}\su{T}$, where
the transverse part ${\mbox{\bf v}}\su{T}$ 
satisfies $\partial_i v\su{T}_i=0$
and the longitudinal part is of the form
 ${\mbox{\bf v}}\su{L}={\mbox{\boldmath $\nabla$}}
\psi_v$. \eq{71} defines the local Hubble parameter $H$, shear
$\sigma_{ij}$ and vorticity $\omega_{ij}$, this last being given by
\be
\omega_{ij} =
\frac12 (\partial_i v\su{T}_j-\partial_j v\su{T}_i)
\label{203a}\;.\ee
Angular momentum conservation gives
$\omega_{ij}\to a\mtwo$, so $\bfv\su{T}$ decays like $a\mone$
and may be dropped (remember that $\partial_i\equiv
a\mone\partial/\partial x^i$).
 
Taking $\bfv$ to be purely longitudinal, it is determined
by the density perturbation in the following way.
First, take the trace of \eq{71} to learn that
$\del. {\bfv} = 3\delta H$. From
\eqss{25}{90}{123} and the matter domination relation
$Ht=2/3$, it follows that 
\be \del. {\bfv}=-(4\pi G\delta\rho ) t \label{147}\;.\ee
The solution of this equation is
\be {\bfv}=-t\del\psi 
\label{147a} \ee
or
\be v_i(\bfx,t)=- (t/a)\pdif{\psi(\bfx) }{x^i} \label{147b}\ee
where 
\be \psi(\bfx) = - G a\mtwo \int\frac{\delta\rho(\bfx^\prime ,t)}
	{|\bfx^\prime -\bfx|} d^3x^\prime  \label{148} \;.\ee
The factor $a\mtwo$ converts coordinate distances into physical distances.
Since $\psi$ 
is related to the density perturbation by the Newtonian expression,
it is called the peculiar gravitational potential. It is independent of
$t$ because, from \eq{32},  $\delta \rho\propto a^2$.
 
From \eq{31} we see that the peculiar gravitational potential is
related to the spatial curvature perturbation by 
\be \psi= - \frac35 {{\cal R}}(\mbox{final})  \label{149} \;.\ee
From \eqs{147a}{148} the Fourier components of $\bfv$, $\psi$ and $\delta$
are related by 
\bea {\bfv}\sk &=& i\frac{\bfk}{k}\rfrac{aH}{k} \delta\sk 
\label{159}\\
\psi\sk &=& -\frac32 \rfrac{aH}{k}^2 \delta\sk
\label{159a}\;.\eea
 
\subsection{Peculiar velocity in general relativity}
 
The Newtonian picture has to be replaced by general relativity
before matter domination, and even after matter domination on scales
that are not far enough inside
the horizon. To define peculiar velocity in this case one can 
proceed as follows \cite{LL2,brly}. 
Start with the definition
\eq{70} of the velocity gradient, and first assume that
the vorticity is negligible since angular momentum conservation ensures 
that it decays like $(\rho+p)a^{-5}$. Then one
can show \cite{brly} that the velocity gradient perturbation is of the form
\be
\delta u_{ij} =\partial_i  v_j +
\frac 12 \dot h_{ij}
\label{duij}
\ee
where $\bfv$ is a globally defined peculiar velocity
field as in the Newtonian picture and 
the extra term $h_{ij}$ is transverse, $\partial_i h_{ij}=0$, and
traceless, $\delta^{ij}h_{ij}=0$.
The extra term represents the effect of gravitational waves,
and at any epoch after matter domination 
the Newtonian picture is valid on scales well inside the horizon
provided that it is negligible. It is
certainly negligible at the present time
in the region of the universe around us with radius a few 
hundred Mpc, where detailed galaxy surveys are performed,
but it could be significant on bigger scales and at earlier times
and so contribute to 
the cmb anisotropy \cite{brly}. Its presence does not however
spoil \eqst{147}{159a} because it is transverse and traceless.
The conclusion, as advertized, is that the Newtonian equations of the last 
subsection apply to the peculiar velocity on all scales after matter 
domination. We shall use them in the next section to derive 
the Sachs-Wolfe effect.
 
One can avoid dropping the vorticity by proceeding as follows \cite{brly}.
First, a transverse velocity $\bfv^T$ may be {\it defined} by \eq{203a}. 
The worldlines with velocity $-\bfv^T$ relative to the comoving worldlines
have no vorticity, and comoving hypersurfaces are defined to be orthogonal to 
them (there are no hypersurfaces orthogonal to worldlines with nonzero 
vorticity). The velocity gradient $\delta u_{ij}$ receives an extra 
contribution $\frac12 (\partial_i v\su{T}_j - \partial_j v\su{T}_i)+
\frac12 (\partial_i w\su{T}_j + \partial_j w\su{T}_i)$ where $\mbox{\bf 
w}\su{T}= \left[1+6\left(1+\frac{p}{\rho} \right)
\left(\frac{aH}{k}\right)^2 \right] {\mbox{\bf v}}\su{T}$.
For scales well inside the horizon $\bfw\su{T}$ is negligible and the 
Newtonian result is recovered.
 
These results demonstrated, within the fluid flow approach,
that the perturbations decouple into three modes. One is associated
with gravitational waves, another with the vorticity of the velocity 
flow, and the third with the density perturbation that is our main
concern.
 
\section{The cmb anisotropy}
 
The first detection by COBE of the intrinsic anisotropy
of the cmb \cite{smet}, announced in 1992, was surely the
most important advance in observational cosmology for a long 
time. The COBE satellite
explores large angular scales, corresponding to linear scales of 
order the particle horizon distance $2H_0\mone\sim 10^4\Mpc$,
and if the standard model is correct it
provides a direct measurement of the primordial 
curvature perturbation. More recently other observations have
detected smaller scale cmb anisotropy,
which in due course will probe the finer details of the 
standard model and help to prove or disprove it
 \cite{scottrev,scottwhite}.
 
The anisotropy of the cmb is defined by giving the 
variation of its intensity with direction, at fixed wavelength.
This variation is usually specified by giving the equivalent variation
in the temperature of a blackbody distribution.\footnote
{In principle the anisotropy depends on the wavelength but 
the dependence is not significant.}
Denoting the direction of 
observation by a unit vector ${\mbox{\bf e}}$, the anisotropy may be expanded 
into multipoles
\be \frac{\Delta T(\bfe)}{T}=\bfw.\bfe+\sum_{l=2}^{\infty} 
	\sum_{m=-l}^{+l} a_l^m Y_l^m (\bfe) \label{mult} \;.\ee
 
The dipole term $\bfw.\bfe$ is well measured, and is the 
Doppler shift caused by our velocity $\bfw$ relative to the rest frame 
of the cmb. Unless otherwise stated, $\Delta T$ will 
denote only the intrinsic, non-dipole contribution from now on. 
 
A feature in the sky of angular size $\theta$ radians is dominated
by multipoles of order $l\sim 1/\theta$.
(This is  analogous to the relation $R\sim 1/k$ between the linear size of a 
feature and the wavenumbers that dominate its Fourier expansion.) Translating 
to degrees we have the following relation between $l$ and the angular scale 
\be \frac\theta{1^0}\simeq \frac{60}{l} 
\label{227}\;.\ee
The cmb 
originates at high redshift, and therefore on a comoving sphere whose 
present distance is close to the particle horizon at $2H_0\mone=6000
h\mone\Mpc$. Its anisotropy also originates at comparable distances.\footnote
{Except for the Sunyaev-Zel'{}dovich effect, which is caused 
by galaxy clusters. It is significant only  on arcminute 
scales, and will be ignored here.}
A feature with angular size $\theta$ therefore corresponds to a 
linear scale 
\be
R\simeq 2H_0\mone\frac{\theta}{1\mbox{\,radian}}=100 h\mone\Mpc
\frac{\theta}{1^0}
\;.\ee
Equivalently, the $l$th multipole of the cmb explores the scale
\be
R\simeq 2H_0\mone/l=6000h\mone\Mpc/l
\;.\ee
 
Present observations have resolution of order 1 degree, so they
give information on scales
$100\Mpc \lsim k\mone \lsim 10^4\Mpc$, and the lower limit will drop
to around $10\Mpc$ when the angular resolution is improved.
There is no significant anisotropy on smaller scales, because it is 
washed out by the finite thickness $\simeq 7h\mone\Mpc$ of the last 
scattering surface.
 
\subsection{The spectrum of the cmb anisotropy}
 
The value of each multipole $a_l^m$ depends on the observer's
position. For a  random choice of position there is some probability 
distribution, with zero mean, and some variance
$\left\langle |a_l^m|^2 \right\rangle\sub{position}$
which is the expected value of $|a_l^m|^2$ for a randomly placed 
observer. The variance is independent of $m$ because no direction is space is 
preferred. By analogy with \eq{4}, one may define the spectrum
$\calp_T(l)$ of the cmb anisotropy by
\be
\calp_T(l)\equiv \frac{(2l+1)l}{4\pi}
 \left\langle |a_l^m|^2 \right\rangle\sub{position} 
\;.\ee
The normalization is chosen so that 
the squared anisotropy averaged over both direction in the sky and
position of the observer is
\be \left\langle\left\langle \rfrac{\Delta T}{T}^2 \right\rangle 
\sub{sky}	\right\rangle\sub{position}
	=\sum^\infty_2 
\calp_T(l)/l
\label{4last} 
\label{226}\;.\ee
 
The cmb anisotropy is well described by linear theory,
so the multipoles at a given position depend linearly on
the Fourier components of the initial curvature perturbation
$\calr$.
According to the standard model $\calr$ is Gaussian, which means 
essentially that the phases of its Fourier components are random.
As a result the spectrum $\calp_T(l)$ depends linearly
on the spectrum $\calp_\calr(k)$ of the initial curvature perturbation,
so their is a `photon transfer function' ${\cal T}_l^2(k)$ such that
$\calp_T(l)=\int^\infty_0{\cal T}_l^2(k)\calp_\calr(k)
\diff k$. Unless there is early re-ionisation, which we shall see 
is unlikely in viable versions of the standard model,
the transfer function is determined by
the amount and nature of the dark matter, through
the same physical processes that we listed already.
The only difference is that, except at the bottom 
end of the relevant range of scales 
$10\Mpc \lsim k\mone \lsim 10^4\Mpc$,
it is not sensitive to the nature of the 
dark matter.
 
The spectrum $\calp_T(l)$ gives $\langle|a_l^m|^2\rangle\sub{position}$,
but we can only measure $|a_l^m|^2$ at our position. The best guess is that 
it is equal to $\langle|a_l^m|^2\rangle\sub{position}$, but
one can also calculate the variance of this guess, which is called the 
{\it cosmic variance}. Because $\calr$ is Gaussian, the real and imaginary 
part of each multipole has an independent Gaussian distribution
and as a result the cosmic variance of $\sum_m|a_l^m|^2$
is only $2/(2l+1)$ times its expected value. 
However, typical present observations suffer from an associated effect, {\em 
sample variance} \cite{sampvar}, which arises 
because they cover only part of the sky 
preventing measurement of the full set of $a_l^m$. In 
essence, cosmic variance is the part of the sample variance which remains 
even once the full sky is observed, due to the recognition that complete 
determination of the spectrum would involve measurement on an ensemble of 
microwave skies rather than on the single one which is accessible.
 
A prediction for the spectrum is shown in Figure 1,
and a qualitatively similar result is obtained in any viable
version of the standard model.
 
\subsection{The large scale cmb anisotropy (Sachs-Wolfe effect)}
 
The simplest regime is the flat 
part of the spectrum corresponding to $l\lsim 30$ or angular scales
bigger than a few degrees.
If the standard model is correct the anisotropy 
in this regime probes directly the initial curvature perturbation,
being caused essentially by the variations in the peculiar gravitational
potential encountered by the cmb on its journey from the last scattering 
surface. This effect is called the Sachs-Wolfe effect, and we calculate 
it now for the case of critical density and zero cosmological constant
following the approach of \cite{LL2,brly}.
The generalisation to the other cases 
\cite{lambda,both,lowomega}
is completely straightforward,
except for the lowest few multipoles where the effect of spatial 
curvature may become important \cite{lystomega}.
 
It is convenient to separate the anisotropy into two parts,
\be
\frac{\Delta T(\bfe)}{T}=\rfrac{\Delta T(\bfe)}{T}\sub{initial}+
\rfrac{\Delta T(\bfe)}{T}\sub{jour}
\label{229}\;.\ee
The first term is the initial anisotropy, measured soon after decoupling
by a set of 
 comoving observers whose detectors point radially outwards and whose
clocks are synchronized (this last condition just says that the 
anisotropy is defined on a comoving hypersurface).
The second term is
the additional anisotropy acquired on the journey towards us,
including the Doppler shifts associated with the peculiar velocities
of the initial comoving observers and of ourselves.
(Our own peculiar velocity 
contributes only to the dipole but for the moment we
are retaining all multipoles.)
 
On the scales in question
the adiabatic initial condition \eq{2} still holds at decoupling.
Since $\rho_r\propto T^4$ it follows that
\be
\rfrac{\Delta T(\bfe)}{T}\sub{initial}\simeq\frac13
\delta(\mbox{\bf x}\sub{em},t\sub{em})
\label{230}
\ee
where $\delta$ is the matter density contrast. Here
$t\sub{em}$ denotes the time of emission (decoupling), and
$\bfx\sub{em}=2H_0\mone \bfe$ denote the comoving coordinates of the
point of emission.
 
To calculated the second term, consider
a photon passing a succession of comoving observers
\cite{LL2,brly}.
 Its trajectory is $a \diff\bfx/\diff t=-{\mbox{\bf e}}$ and 
between nearby observers its Doppler shift is
\be
-\frac{\diff \lambda}{\lambda}=e_i e_j u_{ij} \diff r=-\frac
{\diff \bar a}{\bar a} + e_i e_j \delta u_{ij} dr
\ee
where the first term is due to the average expansion, and the second 
is due to the relative peculiar velocity of the observers. 
Integrating this expression gives 
the redshift of radiation received by us, which was emitted from
a distant comoving source. The unperturbed result is
$\lambda/\lambda\sub{em}=1/a\sub{em}$, and the first order perturbation 
gives
\be 
\rfrac{\Delta T(\bfe)}{T}\sub{jour}=
\int^{x\sub{em}}_0 e_i e_j \delta u_{ij}({\mbox{\bf x}},t) a(t) \diff x
\label{delt}
\;.\ee
The integration is along the 
photon trajectory
\be 
x(t)=\int^{t_0}_t \frac{\diff t}{a} = 3
\left( \frac {t_0}{a_0}- \frac ta \right) \label{xtee}
 \;.\ee
Neglecting any gravitational wave contribution, $\delta u_{ij}$
is the gradient of the peculiar velocity, \eq{9a}, which is given in 
terms of the peculiar gravitational potential by \eq{147a}.
Using \eq{xtee} and integrating by parts one finds
\be
\rfrac{\Delta T(\bfe)}{T}\sub{jour}=
{\mbox{\bf e}}\cdot[{\mbox{\bf v}}(0,t_0)
-{\mbox{\bf v}}({\mbox{\bf x}}\sub{em},t\sub{em})]
+\frac 13 [\psi({\mbox{\bf x}}\sub{em} )-\psi( 0)]
\label{236} 
\;.\ee
The term $\bfe.\bfv(
0,t_0)$ is the Doppler shift corresponding to our peculiar velocity
(in linear theory). It contributes only to
the dipole and will be discounted from now on.\footnote
{By manipulating this expression it
can be shown that in linear theory the dipole is precisely 
the Doppler shift due to our motion relative to the average
motion of everything inside the sphere that emitted the cmb \cite{brly}.}
The term $-\bfe.\bfv(\bfx\sub{em},
t\sub{em})$ is the Doppler shift corresponding to the peculiar
velocity of the initial observer. 
The  remaining term therefore gives the
anisotropy acquired on the journey towards us
{\em if the initial and final anisotropies
are defined by observers with zero peculiar velocity instead of by 
comoving observers}. 
It involves the gravitational potential $\psi$ and can be thought of as 
being due to gravity, though we derived it by considering a sequence of 
Doppler shifts. The term $-\psi(0)/3$ can be dropped
since it does not actually contribute to the anisotropy,
leaving only the term $\psi(\bfx\sub{em})/3$.
 
\eq{236} is exact in linear theory,
but we are considering only
scales that are far outside the horizon when the cmb is emitted.
On these scales \eqs{159}{159a} show that 
\be
|\psi\sk|\gg|\delta\sk|\gg|\bfv\sk|
\label{233}\;.\ee
As a result we can drop the initial Doppler shift, and we can also drop
the initial anisotropy \eq{230}.  This gives the
Sachs-Wolfe formula 
\be 
\frac{\Delta T(\bfe)}{T}
=\frac13 \psi({\mbox{\bf x}}\sub{em} ) 
\label{424} 
\;.\ee
It gives the large scale cmb anisotropy in terms of the peculiar 
gravitational potential, which is related to the curvature 
perturbation by $\psi=-(3/5)\calr$. On the large scales that we are 
considering $\calr$ retains its initial value.
 
\subsubsection*{The Sachs-Wolfe spectrum}
 
Inserting the Fourier expansion of $\calr$ and projecting out 
the multipoles one finds the spectrum 
\cite{peebles82}
\be
\calp_T(l)= \frac{(2l+1)l}{25} 
\int_0^{\infty} \frac{{\rm d}k}{k} \, j_l^2
        ( 2k/H_0 ) \; \calp_\calr(k)
\ee
where $j_l$ is the spherical Bessel function
and $\calp_\calr$ is the spectrum of the initial curvature perturbation.
For a power-law spectrum $\calp_\calr(k)= 
(2k/H_0)^{n-1}\calp_\calr(H_0/2) $ 
one finds
\be
\calp_T(l)=\frac1{50} \frac{2l+1}{l+1}
\left[\frac{\sqrt\pi}{2}l(l+1)
\frac{\Gamma((3-n)/2)}{\Gamma((4-n)/2)}
\frac{\Gamma(l+(n-1)/2)}{\Gamma(l+(5-n)/2)}
\right] \calp_\calr(H_0/2)
\label{422} \;.\ee
For $n=1$ the square bracket is equal to 1.
 
\subsubsection*{The COBE measurement of the cmb anisotropy}
 
The COBE observations provide (after considerable analysis) 
estimates of the multipoles $a_l^m$ 
in the range $2\leq l \lsim 30$. In this regime
the Sachs-Wolfe formula applies
leading to \eq{422}, and the
multipoles are found to be consistent with 
this expression
which is a highly non-trivial test of the standard model.
From \eq{422} one can 
estimate the spectral index
$n$ and the normalization $\calp_\calr(H_0/2)$.
Instead of the latter
one can specify any of the quantities $\calp_T(l)$, which define 
the expected squared multipole. The usual quantity is
the quadrupole, defined by
\be
Q\sub{rms-PS}^2\equiv 2T^2\calp_T(2)
\;.\ee
From \eq{226}, the
factor $2T^2$ makes $Q^2\sub{rms-PS}$ the quadrupole contribution to 
the expected mean square of $\Delta T$.
 
The latest analysis of the COBE data \cite{gorski} finds that a
good fit can be obtained with the spectral index anywhere in the 
range $n\simeq 0.6$ to $1.4$. At fixed $n$ the normalization is however well 
determined, and in particular if $n=1$ then
\be
Q\sub{rms-PS}=(19.9\pm 1.6) \mbox{$\mu$K}
\;,\ee
which corresponds to $\calp_\calr\half=5.8\times10^{-5}$.
 
The best fit value of $Q\sub{rms-PS}$ depends significantly on $n$, but 
according to the analysis of \cite{gorski} 
it becomes practically $n$ independent if 
the 9th multipole is used instead. This fixes the normalization 
for all $n$.
 
\subsection{The smaller scale cmb anisotropy}
 
Once the normalisation has been determined,
the smaller scale anisotropy
can be calculated in a given version of the standard model. It is 
practically independent of the nature of the dark matter
except on very small scales, and barring the possibility of early 
re-ionisation is therefore determined by
$n$, $h$, the matter density $\Omega_m$ and any cosmological constant
contribution $\Omega_\Lambda$. In the (critical density)
MDM model that we are 
focussing on it therefore depends only on $n$ and $h$.
 
As accurate measurements of the smaller scale cmb 
anisotropy are only beginning to be available
firm conclusions are not yet possible, but this
situation will change in the near future.
 
As shown in Figure 1,
the predicted spectrum exhibits 
a series of peaks and troughs as one goes down in scale,
which are usually called `Doppler peaks'.
As an approximation, they may be regarded as coming 
from the standing-wave oscillation of the 
baryon-photon fluid density perturbation,
between horizon entry and decoupling. 
The frequency of the oscillations goes up as the scale goes down,
so the magnitude of the perturbation at decoupling
is an oscillating function of scale. It
causes both an initial anisotropy
$\Delta T\sub{initial}$ as seen by a comoving observer, 
and an
additional anisotropy $\Delta T\sub{jour}$
corresponding to the peculiar velocity of this observer.
 
Re-ionisation before $z=50$ or so would markedly reduce the height
of the peaks, but as we discuss at the end of the next section such an 
early epoch is not predicted by viable versions of the standard model.
The observations seem to be seeing the peaks at more or less the 
expected height \cite{scottwhite} in weak confirmation of this
prediction, but a quantitative analysis is not yet possible.
 
\section{Theory and observation}
 
Now we return to
the density perturbation, and see what constraints are 
placed on it by observations of galaxies and clusters. 
We focus on $\sigma(R,z)$, the {\em rms} of the linearly evolved
smoothed density contrast evaluated at an epoch corresponding to
redshift $z$. As we noted earlier, apart from small corrections if a hot 
dark matter component is present it is proportional to $(1+z)
\mone$, so it is enough to consider the present value
\be
\sigma_0(R)\equiv (1+z)\sigma(R,z)
\;.\ee
Practically all of the data 
can be presented in terms of $\sigma_0(R)$,
which makes it preferable to the 
more widely used power spectrum $\calp_0(k)$.
(A subscript 0 will always denote 
the present linearly evolved quantity.)
 
Some data points and theoretical curves 
are shown in Figures 2a and 2b, the latter being a close up of the former. 
The curves are all normalized to the COBE cmb anisotropy as described
earlier. The full line corresponds to the canonical CDM model
($n=1$, $h=0.5$ and $\Omega_\nu=0$), using 
the CDM transfer function of reference \cite{bobqaisar}
and taking $\Omega_B=0.05$ because of the nucleosynthesis relation
$\Omega_Bh^2=.013\pm.002$. The other lines 
show the effect of 
varying one at a time the parameters $n$, $h$ and $\Omega_\nu$.
To vary $\Omega_\nu$ at fixed $h=0.5$ 
we have used the parameterisation of the transfer function in
reference \cite{bobqaisar}, and to 
vary $h$ at $\Omega_\nu=0$
we have used the approximation that the CDM transfer function 
depends only on $kh\mtwo$. (On small scales and in the 
presence of hot dark matter highly accurate transfer functions
have yet to be published but these are probably adequate for the purpose 
at hand.)
 
Since the curves are rather similar 
and vary over a couple of orders of 
magnitude, the difference between them is best exhibited by normalising 
them all to a single model, which we take to be
the canonical CDM model. The result is shown in Figure 3. 
In what follows the prediction of the canonical CDM
will be denoted by $\sigma_0\su{canon}(R)$.
 
To the extent that all of our COBE normalized curves come close to crossing
at $R\simeq 4000h\mone\Mpc$,
the COBE data can be regarded as an observational point
for $\sigma_0(4000h\mone\Mpc)$ and for convenience
this is shown in the Figures. Its error bar is significant because it shows 
how much
the theoretical curves can be moved up or down, but we
emphasise that the normalization of each curve is 
determined directly from the cmb
anisotropy as described above, not by fitting to this point.
 
The rest of the data points concern galaxies and clusters,
as we now discuss.
 
\subsection{The distribution of galaxies and galaxy clusters}
 
The most extensive observations concern the distribution of galaxies and
clusters, and provide information about $\sigma_0(R)$ on scales
$10\Mpc\lsim R\lsim 100\Mpc$. They also provide information about
the  correlation function
$\xi_0(R)$. These quantities are related to the spectrum by
\bea
\sigma_0^2(R) &=& \int^\infty_0 W^2(kR) \calp_0(k) \diff k/k \label{sigma} 
\\
\xi_0(R) &=& \int^\infty_0 W(kR) \calp_0(k) \diff k/k \label{xi}
\eea
where $W$ is the `top hat' window function, and $\xi_0(R)$ is taken
to be the volume averaged quantity \cite{peadodd}.
A subscript 0 always denotes the present {\em linearly evolved}
quantity.
 
The number density contrasts are known fairly well for
IRAS galaxies, optical galaxies, radio galaxies and Abell galaxy 
clusters, out to a distance of several hundred Mpc.
Within the rather large observational errors 
they are consistent with the biasing hypothesis,
\be
\frac{\delta_I(\bfx)}{b_I}\simeq 
\frac{\delta_O(\bfx)}{b_O} \simeq
\frac{\delta_R(\bfx)}{b_R} \simeq
\frac{\delta_A(\bfx)}{b_A} \simeq
\delta(\bfx)
\ee
where $\delta_I$ etc.~are the number density contrasts,
$\delta$ is the matter density contrast and
$b_I$ etc.~are scale independent bias factors close to unity.
The subscripts I, O, R and A refer to the four classes of objects 
mentioned.
One can use the data to estimate
the bias factors and hence the linearly evolved smoothed density 
contrast, on scales $1\Mpc\lsim R\lsim 100\Mpc$.
(The determination of the overall normalization comes from 
nonlinear corrections, including redshift space to real space 
corrections.)
 
From a number of possibilities, we have chosen 
the analysis of Peacock and Dodds \cite{peadodd},
which combines a variety of data sets.
These authors find $b_I:b_0:b_R:b_A
=1:1.3:1.9:4.5$ for the ratios of the bias factors,
and $b_I=1.0\pm0.2$ for the overall normalization.
Knowing the bias factors, the 
analyses of various groups gives estimates of $\calp_0(k)$,
$\sigma_0(R)$ and $\xi_0(R)$. Peacock and Dodds convert the last two 
into estimates
of $\calp_0(k)$ using the prescription\footnote
{Although this prescription relating $\sigma_0(R)$ to $\calp_0(k)$ is 
adequate on scales $\gsim 10\Mpc$, it is too dependent on the
shape of the transfer function to be useful on smaller scales.
As we shall see, data on such scales directly constrain 
$\sigma_0(R)$, which is our main reason for focusing on that quantity
rather than on $\calp_0(k)$ or $\xi_0(R)$.}
\bea
\sigma_0(R) &=& \calp_0\half(k_R) \label{sigpd} \\
\xi_0(R) &=& \calp_0\half(\sqrt 2 k_R) \label{xipd}
\eea
where 
\be
k_R=\left[\frac12\Gamma\left(\frac{m+3}{2}\right)\right]^{1/
(m+3)}\frac{\sqrt5}{R} \label{kr}
\ee
and $m\equiv (k/\calp_0)(\diff \calp_0/\diff k)$ is the effective 
spectral index
evaluated in the canonical CDM model. 
These formulae are obtained by taking $m$ constant, and using the 
approximation 
\be
W(kR)=\exp(-k^2R^2/10)
\ee
which is exact for $kR\ll1$.
We have used this same prescription to
convert the Peacock-Dodds estimates of $\calp_0(k)$ 
into estimates of $\sigma_0(R)$, and 
the results are shown as open squares in
Figures 2 and 3. For clarity a point with particularly large error bars, 
which lies between the two largest scale points shown, has been omitted.
 
Most of the uncertainty in this determination of $\sigma_0(R)$ comes
from the uncertainty in the {\em overall} normalization,
defined by $b_I=1.0\pm0.2$. 
If all of the raw data used in the determination referred directly
to real space number densities, 
$\sigma_0$ would be proportional to $b_I\mone$ times the 
raw data,
so its uncertainty would be the same as for $b_I$, namely $20\%$.
This is the case for the results of the APM survey,
which is a very important part of the total input, but as
Peacock and Dodds note the rest of the input is based on
redshift space number densities. In the linear regime the determination of 
$\sigma_0$ from raw data on a given type of object in redshift space
is proportional to $[b_I(1+.66rb_I\mone+.20r^2b_I\mtwo)]\mone
$, where $r\equiv b_I/b_N$.
For clusters $r=4.5$ so that $\sigma_0$ is more or less 
proportional to $b_I\mone$
as for real space data, but for the three types of galaxy $r=1.0$,
$1.3$ and $1.9$ and $\sigma_0$ is roughly proportional to $b_I^{-5/7}$.
Thus, galaxy data in redshift space in the linear regime
gives an uncertainty in $\sigma_0(R)$  of order $(5/7)\times20\%
=14\%$. To be on the safe side we
work with $20\%$, noting that at least in the linear regime 
the true uncertainty is between $14\%$ and $20\%$.
 
With $b_I$ fixed, one obtains comparatively 
small error bars, which are indicated by the 
inner error bars in Figures 2 and 3. As an aid to clarity,
in Figure 3 we have connected the ends of these error bars
by straight lines to form a box.
Within the box the points can be moved up and down more or less 
independently, and the whole box can be moved up or down by
$20\%$ as indicated by the outer error bars.
Both these types of error are the result of statistical fitting, and
presumably correspond to something like a 1-$\sigma$ level
(with the caveat that  the outside error bars might be reduced 
somewhat as discussed above).
 
\subsection{The bulk flow}
 
The smoothed peculiar velocity field $\bfv(R,\bfx)$ is called the
bulk flow, and on scales above
$10\Mpc$ it is described by linear theory.
It is therefore the gradient of a potential, and can 
in principle be constructed from its 
radial component \cite{dekelrev}. The latter is directly observed 
from the redshift-to-distance ratios of
galaxies, in the region around us with radius a few hundred Mpc.
 
Unlike the density contrast, the bulk flow can reasonably be 
assumed to be the same as that  of the underlying matter at least on 
large scales, which in principle allows one to dispense with the biasing 
hypothesis.
In practice one still needs the 
hypothesis at present if the data are to yield really powerful results,
which are obtained by 
comparing the density field obtained from the bulk flow with that obtained 
from galaxy surveys. A recent study \cite{dekelet} concludes that at 95\% 
confidence level $b_I=0.7^{+0.6}_{-0.2}$.
 
Although the bulk flow alone does not yet give very powerful results, it 
is not completely useless. The standard way of utilising 
it is to look at the magnitude $v(R,0)$ of the bulk flow
at our position, or in other words at the
average peculiar velocity 
in the sphere around us of radius $R$.\footnote
{There are several variants of this procedure, which correspond to using 
window functions different from a top hat. In what follows we are actually 
using the variant described in \cite{LL2}.}
From \eq{159} the expected value of $v^2(R,0)$ is
\be
\sigma_v^2(R)=\int^\infty_0 \rfrac{H_0}{k}^2 W^2(kR) \calp_0(k) 
\frac{\diff k}{k}
\label{70a}
\;.\ee
Evaluated at random positions, each of the
three components of the bulk flow has a Gaussian 
distribution with one third of this variance, so at
1-$\sigma$ level one expects that 
\be
\frac{v(R,0)}{\sigma_v(R)}=
1^{+32\%}_{-52\%}
\label{71a}
\;.\ee
Observational values of $v(R,0)$ are available only 
for $R$ of order tens of Mpc, and nothing is lost 
by considering just $R=40h\mone\Mpc$. 
The canonical CDM model gives \cite{LL1} $\sigma_v\su{canon}(40h\mone\Mpc
)=460\km\sunit\mone$,
and the observational value is $v\sub{obs}(40h\mone\Mpc
,0)=(400\pm15\%)\km\sunit\mone$.
This error is swamped by the `cosmic variance' 
given by \eq{71a}, and ignoring it
we deduce that 
\be
\frac{\sigma_v(40h\mone\Mpc)}{\sigma\su{canon}_v(40h\mone\Mpc)}
=(400/460)^{+92\%}
_{-24\%}
\ee
where the error is just the inverse of that in \eq{71a}.
 
Compared with \eq{38} for $\sigma_0(R)$, \eq{70a} contains
an extra factor $(H_0/k)^2$ and as a result
$\sigma_v(40h\mone\Mpc)$ probes about the same scale $k\mone$
as $\sigma_0(90h\mone\Mpc)$. We conclude that the above ratio applies
to this latter quantity, which leads to the estimate
$\sigma_0(90h\mone\Mpc)=.044^{+.040}_{-.010}$  shown in Figures
2 and 3.
 
\subsection{Abundances}
 
The rest of our observational points invoke the
well known Press-Schechter estimate \cite{LL2}  for 
the contribution to the mass density of objects with mass $>M$,
\be
\Omega(>M(R),z)=\mbox{erfc}\rfrac{\delta_c}{\sqrt 2\sigma(R,z)}
\label{275}\ee
where $\delta_c=1.7$ and $M(R)$ is the mass in a comoving sphere
of present radius $R$, given by \eq{mass}.
Except for a more or less unmotivated factor 2 the right hand side is 
just the fraction of space occupied by regions of space 
where the linearly evolved smoothed density contrast 
exceeds $\delta_c$, so roughly 
speaking the Press-Schechter estimate simply states that such regions 
are occupied by objects with mass bigger than $M$.
It is only supposed to be valid as long as these regions are rare,
corresponding to $\Omega(>M,z)\ll 1$. In this regime
$\sigma(R,z)\ll1$, and linear evolution should be valid.
 
The value $\delta_c=1.7$ is motivated by a 
spherical collapse model, and N-body simulations suggest that the
Press-Schechter estimate is roughly correct, if
$\delta_c$ is within ten percent or so of this value
\cite{lacey,mabert2}.\footnote
{Somewhat smaller values are found with a Gaussian instead of a top-hat 
smoothing but we are using the latter.}
 
In some cases the observational result is given directly as 
a value of $\Omega(>M)$, but in others it is given as a number density 
$n(>M)$. In the latter case we obtain an observational value of
$\Omega(>M)$
by setting $\rho(>M)\simeq Mn(>M)$
which gives
\be
\Omega(>M)=3.60 h \frac{M}{10^{12}\msun} \frac{n(>M)}
{(h\mthree \Mpc)\mthree}
\label{81}
\;.\ee
This overestimates $\Omega(>M)$ and therefore $\sigma(R)$,
but in practice the mass of the observed objects is 
too uncertain for this to matter very much.
 
An alternative is to
obtain a theoretical prediction for $n(>M)$ 
from the Press-Schechter formula. This more complicated 
procedure has been used by many authors
(starting with the originators of the formula), but 
because of the uncertainty in the mass there is generally no significant 
advantage in using it.
Some relevant comparisons are cited below.
 
\subsubsection*{Galaxy clusters}
 
We first look at
galaxy clusters. The most useful are those 
of `richness class $R>1$', which are the heaviest bound objects.
Their present number density is known to be about
$n\simeq 8\times 10^{-6} h^3\Mpc\mthree$ and their mass 
$M\sub{clus}$ is roughly $10^{15}\msun$. The uncertainty in the mass
is the limiting factor in the present context.
Let us first set it equal to $1.2\times 10^{15}\msun$ which corresponds
(with $h=.5$) to $R=8h\mone\Mpc$. Then \eqs{275}{81} give
$\sigma_0(8h\mone\Mpc)=0.71$.
Lowering $M\sub{clus}$ by a factor 2 gives
$\sigma_0(6.4h\mone\Mpc)=0.64$,
and raising it by a factor 2 gives 
$\sigma_0(10.1h\mone\Mpc)=0.80$.
These three values are indicated in Figures 2a and 2b
by a central point and a slanting `error bar'.
Because $\sigma_0\su{canon}(R)$ is strongly varying 
the corresponding uncertainty in
$\sigma_0(R)/\sigma_0\su{canon}(R)$ is
quite big, namely $+43\%$ and $-27\%$. 
Compared with this uncertainty
the range of scales $R=6.4h\mone$ to $10h\mone\Mpc$ is negligible 
(essentially because all the spectra we consider have broadly similar 
shapes),
and we have ignored the horizontal error when plotting Figure 3.
 
Our conclusion is that if $R>1$ clusters have mass 
$M=10^{15}\msun$ within a factor 2, the normalization of the
standard model of structure formation is 
$\sigma_0
(8h\mone\Mpc)=0.71^{+43\%}_{-27\%}= 0.5$ to $1.0$.
Estimates of the normalization in a similar spirit
have been made by previous authors.
A recent and typical one \cite{whitecluster} 
quoted a normalization $\sigma(8h\mone\Mpc)\simeq 0.57$,
but did not quantify the effect of the uncertainty 
in the cluster mass.
 
One can in principle apply the same technique to clusters observed at 
high redshift. A treatment of $z=0.3$ observations 
\cite{carlberg93} which estimates the effect of both
mass and number density uncertainty gives
$\sigma_0(8h\mone\Mpc)=0.6$ to $0.9$, in excellent agreement with 
our result.
High redshift observations of clusters
are likely to become increasingly important in the future.
 
\subsubsection*{Quasars and damped Lyman alpha systems}
 
The same technique can be applied to quasars and damped Lyman alpha 
systems observed at redshift 3 to 4.
Their masses are thought to be roughly in the range $10^{11}$ to
$10^{13}\msun$ and on these scales the Press-Schechter 
formula gives $\Omega(>M,z)\ll 1$, indicating that linear theory
applies and the formula is valid.
An observational value of the quasar fraction $\Omega\sub{quas}
(>M,z)$ or the damped Lyman alpha system fraction
$\Omega\sub{lyalph}(>M,z)$ therefore provides a lower limit on
$\Omega(>M,z)$, and on $\sigma_0(R)$.\footnote
{For the purpose of this article `quasar mass' is taken to
mean the total mass of the 
object concerned, though the term 
`quasar' usually refers only 
to the region around the central black hole.}
$\Omega\sub{quas}(>M,z)$ can be estimated from observation 
out to $z=4$, in the range 
roughly $10^{11}\msun$ to $10^{13}\msun$, if one knows
the baryon fraction that forms the central black hole of each quasar
and some other astrophysical parameters.
Haehnelt \cite{haehnelt} 
takes $1\%$ as a reasonable estimate of the 
fraction, and using reasonable values for the other parameters 
estimates that $\sigma_0(M=10^{13} \msun)>1.30$.
Multiplying the baryon fraction by 10 (or making equivalent changes in 
the other parameters) reduces this to
$\sigma_0(M=10^{13} \msun)>1.25$, which is shown in figures 2 and 
3.\footnote
{The weak dependence on the baryon fraction arises because
$\Omega\sub{quas}(>10^{13}\msun,3)$ is tiny, so that the prediction for it 
is an exponentially decreasing function of $\sigma_0$.}
 One sees that this is not very constraining.
He also gives bounds for lower masses, down to $M=10^{11}\msun$
but they are even less restrictive. 
(A more recent estimate \cite{mabert2} using newer data 
suggests that a better constraint might now be possible, but a detailed 
analysis has not yet been done.)
 
More restrictive bounds \cite{lymanalpha} come from damped Lyman alpha
systems. The data indicate
\cite{wolfe} that the {\em baryons}
in such systems account for $\Omega\sub{lyalph}(>M,(3.0\pm0.5))=
(0.0029\pm 0.0006)h\mone$. Let us take the lower 
bound of the redshift and $\Omega$ to obtain a lower bound
on $\sigma_0$ (we set $h=0.5$ as its uncertainty is 
not significant). Then
if the baryon fraction is the same as the universal fraction
$\simeq 0.05$, we deduce that
$\Omega(>M\sub{lyalph},3.0)>0.092$ and
$\sigma_0(R)>4.0$ where $R$ is the scale corresponding
to $M\sub{lyalph}$.
To obtain a reasonably firm 
bound let us divide this result by 10 (corresponding, say, 
to assuming that the 
baryons are concentrated by a factor 10).
This leads to $\sigma_0(R)>2.6$.
 
The corresponding bound on $\sigma_0(R)/\sigma_0\su{canon}(R)$
depends on $M\sub{lyalph}$,
but the dependence is much weaker than it is for clusters
because $\sigma_0\su{canon}(R)$ is varying more slowly
with scale. The bounds displayed in Figures 2 and 3 are for
$M\sub{lyalph}=3\times 10^{11}\msun$, corresponding to
$R=0.50h\mone\Mpc$. Going 
down to $10^{10}\msun$ only lowers $\sigma_0(R)/\sigma_0\su{canon}(R)$
by $17\%$ and going up to $10^{12}\msun$ only  raises it by
$7\%$, the corresponding scales being $R=0.16h\mone\Mpc$
and $R=0.76h\mone\Mpc$. None of these changes is very significant.
 
Our Lyman alpha constraint corresponds to a 
constraint $\Omega_\nu\lsim 0.2$  (with $n=1$ and $h=0.5$).
This is the same as that obtained in most (but not all) previous 
investigations \cite{lymanalpha,silvio},
which have used both the Press-Schechter formula and numerical 
simulations.
 
\subsection{Other observations}
 
The observations that we have considered are the only ones that can 
be related more or less directly to the linearly evolved density
contrast. Other observations can in principle yield information after
comparing them with numerical simulations, but at present 
the uncertainties 
seem to be too big to make the additional information useful.
For example, the pairwise galaxy velocity dispersion 
has been widely compared with numerical simulations
and the received wisdom is that with the COBE normalization the 
prediction is significantly too high (eg. \cite{gelb}). 
But others \cite{zurek} argue that when the observational
\cite{mojing}
and theoretical uncertainties are taken into account the
discrepancy ceases to be significant. 
 
\subsection{Summary of the data on $\sigma_0(R)$}
 
The most accurate data point is the one coming from the COBE 
measured CMB anisotropy, which is 
indicated schematically by the point at $4000h\mone\Mpc$.
That is, of course, the reason why we have used it to normalize the 
theoretical curves.
 
The other points all come from galaxy and cluster observations,
and for clarity we will focus on $\sigma_0(R)/\sigma_0\su{canon}(R)$
as displayed in Figure 3. Most of the points
come from 
the number density contrasts of various types of object, as analysed 
by Peacock and Dodds, which are represented by open squares.
Within the box each point can be moved down more or less independently, 
and the whole box can be moved up about $20\%$ as indicated by
the outer error bars.
(Both of these uncertainties are more or less at 
the 1-$\sigma$ level, though the $20\%$ estimate may be somewhat
high according to this criterion.)
The most striking feature of these points is the slope
of $\sigma_0(R)/\sigma_0\su{canon}(R)$. 
In the regime $10h\mone\Mpc\lsim R\lsim 40h\mone\Mpc$ 
it is positive and rather well 
defined. This positive slope
is seen in both
the galaxy correlation function (in 
many different surveys) and in the cluster correlation function
\cite{dalton}, and in addition it seems to be needed to make 
N-body simulations give the correct {\it normalization}
for the cluster correlation function \cite{dalton}. In other words,
it is rather firmly established. On smaller scales the slope 
is flatter, and there is also a hint of 
flattening on larger scales.
 
What about the overall normalization of these points?
The $20\%$ uncertainty indicated by the outer error bars
corresponds to at at least a 1-$\sigma$
confidence level in the context of the Peacock and Dodds analysis,
and both the upper and lower limits do have 
confirmation from other types of data.
The {\em lower} limit corresponds to the {\em upper} limit of 
Peacock and Dodds' bias factor
determination $b_I=1.0\pm0.2$, which is confirmed by 
the result \cite{dekelet} 
$b_I=0.7^{+0.6}_{-0.2}$ {\it at 95\% confidence level}, that comes
from combining the galaxy distribution with
observations of the bulk flow.
The lower limit of the normalization is also confirmed,
less strongly, by the bulk flow data on the scale
$R\sim 90h\mone\Mpc$. Coming to the upper limit
on the normalisation of $\sigma_0(R)/\sigma_0\su{canon}(R)$,
one sees from Figure 3 that it
is confirmed on the scale $R\sim 10h\mone\Mpc$ by the abundance
of present day rich galaxy clusters if one assumes that their  mass 
is less than $2\times 10^{15}\msun$.
A mass in this range is certainly commonly assigned to these
rich clusters, and indeed a lower (median) mass corresponding to
a lower upper limit on $\sigma_0(8h\mone\Mpc)$ has been advocated
\cite{whitecluster}. Essentially the same upper limit is
confirmed by the abundance of $z=0.3$ clusters as cited earlier.
 
All this is on scales 10 to 100 Mpc. A crucial lower bound
on the normalization on some scale in the decade $0.1$ to $1\Mpc$
is provided by the abundance of damped Lyman alpha systems
at $z=3$. In contrast with the case of galaxy clusters,
the uncertainty about the precise scale matters little here,
because the theoretical curves that one is trying to normalise are 
almost flat. It is noteworthy that this lower limit is already
presaged by the flattening of the slope seen 
in the regime $4h\mone\lsim R\lsim 10h\mone\Mpc$.
 
\subsection{Theory versus observation}
 
The first conclusion to draw from the comparison of theory and 
observation is that the canonical CDM model does not work. As discussed
in the Introduction three fixes are possible. One is an `old age' model 
where $h$ is very low. Another is a `tilted' model where $n$ is 
significantly below 1. The third is to alter the nature or the amount of 
the cold dark matter, and the specific option that we have explored is 
to introduce some fraction $\Omega_\nu$ of hot (neutrino) dark matter 
while keeping the total matter density fixed at the critical value.
 
Figure 3 shows that any of these three options is capable of 
giving an acceptable fit to the data, the required parameters
being $h\simeq0.3$ for the first option, $n\simeq0.7$ for the second and 
$\Omega_\nu\simeq 0.15$ to $0.20$ for the third. 
In each case more extreme
values (smaller $h$ or $n$ or bigger $\Omega_\nu$)
are ruled out be the observational lower bounds on the normalization of 
$\sigma_0(R)$, especially the one coming from damped Lyman alpha systems
(recall that the one we show is supposed to be rather firm).
Less extreme values are ruled out by the 
{\em slope} of $\sigma_0(R)$ 
(and in the case of the MDM option also by
the upper limit on its magnitude) 
in the tens of Mpc range.
 
What can we say about the relative merits of the three options?
As noted earlier, in most (particle physics motivated) models 
of inflation that implement the tilted option 
there is also a gravitational wave contribution whose effect is to lower 
the COBE normalization by $[1+6(1-n)]\mhalf$. 
For $n=0.7$ this factor is $0.6$ which leads to a gross contradiction 
with the data. As a result the tilted option is not viable in 
most models of inflation, the only known exception being
`natural' inflation \cite{natural1,natural2}.
The old age option is also problematical because of the low
Hubble constant that it invokes, which leaves the
MDM option as perhaps the most attractive of the three.
 
So far everything is quite simple, but we have only allowed 
the three parameters to depart from their canonical values
$h=0.5$, $n=1.0$ and $\Omega_\nu=0$ one at a time.
We would like to understand the entire parameter space.
In particular we would  like to
know the observational constraint on $n$ because,
especially in conjunction with gravitational waves,
it is such a useful discriminator between models of inflation.
 
One way forward 
\cite{pogosyan,bobqaisar,boblast}
is to perform a least squares fit, though this is somewhat 
problematical because the damped Lyman alpha systems
provide only a bound, and because 
it is not clear that all of the points coming from galaxy and cluster
correlations are independent. An alternative \cite{mymnras}
is to simply demand that the curves go through selected points,
and have slopes within the limits indicated by the box.
These options will be explored in a future paper \cite{bobnext},
but for the moment we point out that rather definite 
conclusions \cite{capri}
about $n$ can already be drawn just from Figure 3.
 
To start with, note 
that in the regime where we have data,
a change in $h$ is roughly equivalent
to a change in $n$ according to the formula $\Delta h=-\Delta
n$. This can be seen by comparing the $h=0.3$ and $n=0.7$ curves
in Figure 3, especially for the crucial slope.
Accepting this and taking $0.4\lsim h\lsim 0.6$,
one concludes that with critical density CDM requires
$n\simeq 0.6$ to $0.8$. 
 
The crucial point now is that
{\em all} known modifications of the pure, critical density  CDM
hypothesis {\em reduce} the COBE normalized $\sigma_0(R)$,
because CDM maximises the growth of the density perturbation on small 
scales. Thus the {\em lower} limit $n\gsim 0.6$ will hold in 
{\em any} version of the standard model.
If gravitational waves reduce the COBE normalization by
a factor $[1+6(1-n)]\mhalf$, it is not difficult to verify that
the lower limit is tightened to $n\gsim 0.8$
 
What about the upper limit? On scales $R\lsim 40h\mone\Mpc$ or so one 
can clearly cancel an increase in $n$ by increasing the fraction of 
hot dark matter, but at least within the critical density MDM option
this is impossible on larger scales. Assuming that no cancellation is 
possible, the upper bound $n\lsim 1$ indicated by the 
four largest scale galaxy/cluster correlation points shown
in Figure 3 can only be evaded by decreasing $h$,
which (assuming that $h>0.4$) allows
it to go up to $1.1$. A similar upper bound comes from the fact 
that too much hot dark matter gives a 
strong violation of the damped Lyman alpha system bound, which cannot be 
cancelled by increasing $n$ because of the data in the tens of Mpc 
range. The conclusion is that, at least within the MDM
option, there is an upper bound $n\lsim1.1$ or so.
 
\subsubsection*{The epoch of re-ionisation}
 
We end with a brief discussion of estimates of the re-ionisation
epoch $z\sub{ion}$, which are as yet rather preliminary.
 
Observationally, a 
lower limit of order $z\sub{ion}\gsim5$ is well established
\cite{schneider}, and a useful upper limit or even a value
may soon become available from the small scale cmb anisotropy
\cite{tegsilk}. 
 
To estimate the re-ionisation epoch theoretically, one starts
by estimating the fraction $f$
of the mass that has to collapse into galaxies before re-ionisation 
occurs. 
At present the astrophysics is not understood sufficiently well to 
allow a firm calculation of $f$.
Tegmark, Silk and Blanchard \cite{tegsilk} give a `best guess'
$f\simeq4\times10\mthree$ with upper and lower limits
$2\times10^{-5}\lsim f\lsim 0.4$.
 
If $M\sub{min}$
is the mass of the smallest galaxies existing at the
epoch of re-ionisation,
then $\Omega(>M\sub{min},z\sub{ion})=f$,
and the Press-Schechter
estimate gives $z\sub{ion}$ in terms of $\sigma_0(R\sub{min})$
and $f$. We take as a best guess
$M\sub{min}=10^6\msun$ (the same as the minimum mass of present day 
galaxies), while noting that theoretical estimates vary from
$10^5\msun$ to $10^8\msun$ \cite{tegsilk}. The corresponding scale is
$R\sub{min}\simeq 0.01\Mpc$, two orders of magnitude smaller than the 
scales that we have considered so far.
 
If $f\sim 1$ re-ionisation will 
not occur until the epoch of non-linearity
given by $(1+z\sub{nl})=\sigma_0(R\sub{min})$, but if $f$ is
smaller it will be earlier. Tegmark {\it et al}'s 
`best guess' gives $(1+z\sub{ion})\simeq 1.7 \sigma_0(R\sub{min})
$ whereas the lower end of their allowed range gives 
$(1+z\sub{ion})\simeq 2.5 \sigma_0(R\sub{min})$.
The range is not too big, because if $f$ is small it 
is an exponentially decreasing function of $\sigma_0(R\sub{min})$.
 
\begin{table}
\begin{center}
\caption[Table]{Estimates of the re-ionisation epoch $1+z\sub{ion}$}
\begin{tabular}{|c|c|c|c|c|c|}
\hline\hline
$h$ & $n$ & $\Omega_\nu$ & low estimate & best guess & high estimate \\
\hline
0.5 & 1.0 & 0 & 31 & 53 & 78 \\
0.5 & 0.7 & 0 & 9.0 & 15 & 23 \\
0.5 & 1.0 & 0.3 & 3.9 & 7.6 & 12 \\
0.3 & 1.0 & 0 & 6.1 & 10 & 15 \\
\hline\hline
\end{tabular}
\end{center}
\end{table}
 
Tegmark {\it et al} go on to calculate $\sigma_0(0.01\Mpc)
/\sigma_0(8h\mone)$ in various models. Inserting the
values of $\sigma_0(8h\mone)$ following from our COBE normalization 
we deduce the results in Table 1.
The low estimate of $z\sub{ion}$ 
takes it to be the epoch of non-linearity, 
corresponding to $f\sim 1$, the other two
correspond to the values of $f$ mentioned earlier.
In the third row we have allowed for the slower growth of the MDM 
density contrast in the same way as Tegmark {\it et al} (following
\cite{davis}), and in the last row we have used the scaling
$\sigma_0(R)\propto h^2$ that is appropriate in the small
scale regime where $\sigma_0(R)$ is practically independent 
of $R$. Note that the third row is for $30\%$ hot dark matter, 
whereas at most $20\%$ or so is viable. Thus a viable
MDM option will give earlier re-ionisation 
than the other two options.
 
Although the uncertainties are big, it seems clear that
viable versions of the standard model do not lead to very early
re-ionisation. On the other hand a re-ionisation redshift in the low
tens, which might be big enough to be affect 
the cmb anisotropy, is not out of the question.
 
\section{Conclusion}
 
To conclude, we have provided a review of the machinery required in order to 
translate an initial spectrum of density irregularities
into a form amenable 
for fairly direct comparison with a range of observations.
Although some of our discussion has wider application, we
have focussed on the standard model of structure formation,
which invokes an initially Gaussian, adiabatic, scale invariant
density perturbation and more or less cold dark matter.
It is at present the one under the most active consideration, because it 
is the simplest, and also because the required initial density 
perturbation might be generated as a vacuum fluctuation
during inflation.
 
We have gone on to compare the standard model
with presently available data, indicating the relative merits of 
altering the Hubble constant, tilting 
the primordial spectrum and incorporating a component of hot dark matter. All 
of these have been envisaged as ways to remedy the shortcomings of the 
canonical cold dark matter model.
 
Our treatment of the observations breaks new ground 
in that it combines for the first time several relatively 
new observational constraints.
For the normalization of the
spectrum from the COBE measurement of the
cmb anisotropy, we use the most recent determination 
that has been
provided by G\'{o}rski and collaborators. It is 
significantly higher than earlier determinations.
We assess carefully the
uncertainty in the determination of the normalization on the
scale $8h\mone\Mpc$ that comes from the cluster abundance, 
quantifying its dependence on the assumed cluster mass.
We establish a simple and rather firm constraint from damped
Lyman alpha systems at high redshift, which provide a powerful lower 
limit on some scale of order $0.1$ to $1\Mpc$.
Finally, on the basis of an earlier study \cite{tegsilk}
we make a preliminary estimate of the epoch of re-ionisation
in the various models. 
 
Our technique for comparing theory with observation
is to look not at the power spectrum, but at
the dispersion of the density contrast smoothed on a scale $R$. This has the 
great advantage of being closer to what is actually measured, while still 
being simple to calculate for a given theoretical model. 
Particularly when
one normalises theories and observations to a benchmark model 
as in Figure 3, the comparison between 
theory and observation can be very clearly illustrated.
 
While present observations 
are strong enough to exclude the standard CDM model, 
they are not of sufficient quality to select amongst different ways of 
generalising it, particularly when one realises that if tilt or a hot 
component are to be varied then one must certainly also allow $h$ to vary in 
combination. It is clear from Figure 3 that these options are most cleanly 
probed by observations on scales from a few tens 
to a few hundreds of megaparsecs, and we 
should look forward to this region of the spectrum being probed by both small 
scale microwave anisotropy experiments and by larger scale galaxy correlation 
(and peculiar velocity) measurements.
 
\section*{Acknowledgements}
ARL is supported by the Royal Society, and acknowledges use of the Starlink 
computer system at the University of Sussex. It is a pleasure to thank Bob 
Schaefer, Qaisar Shafi and Pedro Viana for many discussions.

\section*{Figure Captions}
 
\noindent
{\em Figure 1.  [NOT AVAILABLE VIA BULLETIN BOARD]}\\
This figure is reproduced from \cite{cretal}, and it shows
predictions for the spectrum $\calp_T(l)$ of the cmb anisotropy.
The full line `S' shows the contribution (normalized to 1 at $l=2$)
of an adiabatic density perturbation with 
$n=0.85$, $h=0.5$, $\Omega_B=0.05$
and critical total density. It was calculated for pure CDM,
but it is insensitive to the nature of the dark matter.
(The dashed line is the prediction with $\Omega_B=0.01$
and the same value for $h$, but note that these quantities are 
actually linked by
the nucleosynthesis relation $\Omega_B h^2=0.013\pm.002$.)
The line `T' shows the gravitational wave contribution,
also normalized to 1 at $l=2$. 
As discussed in the text, it is not yet known whether this
contribution is actually significant.
 
\vspace{40pt}
\noindent
{\em Figures 2a and 2b.}\\
Observational data, shown in comparison to the canonical CDM model plus four 
variants, two with a hot component, one tilted and one with a low Hubble 
constant. All of the models are normalized on large scales to the COBE data 
as described in the text, and this data is represented
schematically by the filled triangle at $Rh\mone\sim 4000\Mpc$.
The rest of the data comprises damped 
Lyman alpha systems and quasars (two lower 
limits on the left hand side), cluster abundance (triangle, with a tilted 
error bar indicating the dependence of the prediction on the assumed cluster 
mass), galaxy correlations (squares, error bars explained in text and in 
figure 3 caption) and bulk velocities (star). 
The second figure is a closeup of the first, which omits the 
COBE point.
 
\vspace{20pt}
\noindent
{\em Figure 3.}\\
This is the same as Figures 2a and 2b except that everything has been divided 
by the prediction of the 
canonical CDM model. As explained in the text the galaxy correlation
points enclosed by the `error box' can be moved up and down 
more or less independently within the 
box, and in addition the whole box can be moved up and down by the 
amount indicated by the outer error bars. With this exception all of the 
error bars are more or less statistically independent, and represent
something like a 1-$\sigma$ confidence level. The lower bounds are 
intended to express a high level of confidence as explained in the text.
 

\begin{thebibliography}{99}
\bibitem{smet} G. F. Smoot {\it et. al.}, {\it Astrophys. J. Lett.}
{\bf 396} (1992) L1.
\bibitem{gorski} K. G\'{o}rski {\em et al}, ``On determining the
spectrum of primordial inhomogeneity from the 
COBE DMR sky maps: II. Results of two year data analysis'',
COBE preprint (1994).
\bibitem{scottrev} M. White, D. Scott and J. Silk,
to appear in {\it Ann. Rev. Astron. Astroph.} (1994).
\bibitem{scottwhite} D. Scott and M. White, ``The existence
of baryons at $z=1000$'', Berkeley preprint (1994).
\bibitem{will}
A. R. Liddle, D. H. Lyth and W. J. Sutherland, {\it Phys. Lett.} 
	{\bf B279} (1992) 244.
\bibitem{LL1} A. R. Liddle and D. H. Lyth, {\it Phys. Lett.} {\bf B291} 
(1992) 391.
\bibitem{LL2} A. R. Liddle and D. H. Lyth, {\it Phys. Rep.} {\bf 231} (1993)
1.
\bibitem{mymnras} A. R. Liddle and D. H. Lyth, {\it Mon. Not. Roy. astr. 
Soc.} {\bf 265} (1993) 379.
\bibitem{berkeley}
A. R. Liddle and D. H. Lyth, {\it Annals of the New York Academy of Sciences}
        {\bf 688} (1993) 653.
\bibitem{capri} D. H. Lyth and A. R. Liddle, to appear in 
{\it Astrophys. Lett. \& Commun.} (1994)
(proceedings of the Capri workshop);
D. H. Lyth, in {\sl Proceedings of the 1993 Trieste Summer School}
(World Scientific Press, Singapore, 1994).
\bibitem{paul} R. L. Davis, H. M. Hodges, G. F. Smoot, P. J. Steinhardt
and M. S. Turner, {\it Phys. Rev. Lett.} {\bf 69} (1992) 1856.
\bibitem{davesal}
D. S. Salopek, {\it Phys. Rev. Lett.} {\bf 69} (1992) 3602.
\bibitem{lowh} J. G. Bartlett {\em et al},
 ``The case for a Hubble constant of $30\km\sunit\mone\Mpc\mone$'',
preprint (1994).
\bibitem{tilt} 
N . Vittorio, S. Matarrese and  S. Lucchin, {\it
Astrophys. J.} {\bf328} (1988) 69;
J. P. Ostriker and Y. Suto, {\it Astrophys. J.} {\bf 348} (1990)
378; J. R. Bond, in {\sl Highlights in Astronomy} vol. 9 
	Proceedings of the IAU Joint discussion, Ed. J. Bergeron 
	(Buenos Aires Gen Ass) (1992);
R. Cen, N. Y. Gnedin, L. A. Kofman and J. P. Ostriker,
{\it Astrophys. J.} {\bf 399} (1992) L11;
D. S. Salopek, {\it Phys. Rev.} {\bf45} (1992) 1139.
\bibitem{natural2}
F. C. Adams, J. R. Bond, K. Freese,
J. A. Frieman and A. V. Olinto, {\it Phys. Rev.} D{\bf 47} (1993)
426.
\bibitem{gelb} J. M. Gelb, B.-A. Gradwohl and J. A. Frieman, {\it Astrophys. 
J. Lett.} {\bf 403} (1993) L5.
\bibitem{mdm}
S. A. Bonometto and  R. Valdarnini, {\it Phys. Lett.} {\bf 103A} (1984) 
369; Q. Shafi and F. W. Stecker, {\it Phys. Rev. Lett.} {\bf 53} (1984) 1292;
L. Z. Fang, S. X. Li and S. P. Xiang, {\it Astron. Astrophys.} {\bf 140},
(1984) 77;
R. Valdarnini and S. A. Bonometto, {\it Astron. Astrophys.} {\bf 146} (1985) 
235;
S. Achilli, F. Occhionero \& R. Scaramella, {\it Astrophys. J.}
	{\bf 299} (1985) 577;
S. Ikeuchi, C. Norman \& Y. Zhan, {\it Astrophys. J.} {\bf 324}, 
	(1988) 33;
R. K. Schaefer, Q. Shafi and F. Stecker, {\it Astrophys. J.} 
	{\bf 347} (1989) 575;
J. Holtzman, {\it Astrophys. J. Supp.} {\bf 71} (1989) 1;
E. L. Wright {\it et al}, {\it Astrophys. J. Lett.} {\bf 396} (1992) L13;
R. K. Schaefer and  Q. Shafi, {\it Nature} {\bf 359} (1992) 199;
A. N. Taylor and  M. Rowan-Robinson, {\it Nature}, {\bf 359} (1992) 396;
T. van Dalen and R. K. Schaefer, {\it Astrophys. J.} {\bf 398} (1992) 33;
J. Holtzman and J. Primack, {\it Astrophys. J.} {\bf 405} (1993) 428;
A. Klypin, J. Holtzman, J. R. Primack, and E. Reg\"{o}s, 
{\it Astrophys. J.} {\bf 416} (1993) 1;
Y. P. Jing, H. J. Mo, G. B\"orner \& L. Z. Fang, to appear, {\it Astron. and 
Astrophys.} (1994);
S. Ghigna {\em et al}, ``Void analysis as a test for dark matter 
models'', preprint (1994);
G. Yepes {\em et al}, ``The angular correlation function of galaxies
in cdm and chdm models'', preprint (1994).
\bibitem{lymanalpha}
K.  Subramanian and T. Padmanabhan, ,
        ``Constraints on the models for structure formation from the 
abundance of damped lyman alpha systems'', IUCAA preprint (1994);
H. J. Mo and J. Miralda-Escude, 
        ``Damped lyman alpha systems and galaxy formation'',
Princeton preprint (1994);
G. Kauffmann and S. Charlot,
        ``Constraint on models of galaxy formation from the evolution of 
damped Lyman alpha absorption systems'', Berkeley preprint (1994);
A. Klypin, S. Borgani, J. Holtzman and J. Primack, Perugia preprint (1994).
\bibitem{davis} 
M.  Davis, F. J. Summers and D. Schlegel, {\it Nature} {\bf 359} (1992) 393.
\bibitem{pogosyan}
D. Yu. Pogosyan and A. A. Starobinsky, {\it Mon. Not. Roy. astr. Soc.} {\bf 
265}  (1993) 507.
\bibitem{bobqaisar} R. K. Schaefer and Q. Shafi, {\it Phys. Rev. D}
{\bf 49} (1994) 4990.
\bibitem{silvio}
S. A. Bonometto, S. Borgani, S. Ghigna, A. Klypin and J. R. Primack,
submitted to {\it Mon. Not. Roy. astr. Soc.} (1993).
\bibitem{boblast} G. Dvali, Q. Shafi and R. Schaefer, ``Large
Scale Structure and Supersymmetric Inflation Without Fine
Tuning'', Pisa preprint (1994).
\bibitem{lambda}
L. A. Kofman, N. Y. Gnedin and N. A. Bahcall {\it
Astrophys. J.} {\bf 413} (1993)  1;
R. Cen, N. Y. Gnedin and J. P. Ostriker, {\it Astrophys. J.}
{\bf 417} (1993) 387;
R. Cen and J. P. Ostriker, ``X-ray clusters in a CDM$+\Lambda$
universe;  a direct, large scale, high resolution, hydrodynamic
simulation'' Princeton preprint (1994);
E. F. Bunn and N. Sugiyama, ``Cosmological-Constant Cold
Dark Matter Models and the COBE Two-Year Sky Maps'',
Berkeley preprint (1994).
\bibitem{both}
N. Sugiyama and J. Silk, ``The imprint of $\Omega$ on the cosmic 
microwave background'', to appear in {\it Phys. Rev. Lett.} (1994).
\bibitem{tegsilkopen}
M. Tegmark and J. Silk, ``Reionization in an open CDM universe:
implications for cosmic microwave background fluctuations'',
Berkeley preprint (1994).
\bibitem{lowomega} 
M. Kamionkowski, D. N. Spergel and N. Sugiyama,
{\it Astrophys. J.} {\bf 426} (1994) L57;
M. Kamionkowski and D. N Spergel, to appear in
{\it Astrophys. J.} (1 September 1994);
B. Ratra and P. J. E. Peebles, ``CDM cosmogony in an open universe'',
Princeton preprint (1994);
M. Kamionkowski, B. Ratra, D. N. Spergel and N. Sugiyama,
``CBR anisotropy in an open inflation, CDM cosmogony'',
Princeton preprint (1994);
B. Ratra, `CDM cosmogony in an open universe'', Princeton preprint
(1994);
P. Coles and G. Ellis, ``The case for an open universe'', QMW preprint
(1994).
\bibitem{lifs}
E. M. Lifshitz, {\it J. Phys. (Moscow)} {\bf 10} (1946) 116.
\bibitem{peebles80}
P. J. E. Peebles, {\sl The Large Scale Structure of the Universe }
(Princeton University Press, 1980).
\bibitem{synch}
G. Efstathiou, in ``The Physics of the Early Universe'', eds
	Heavens, A., Peacock, J. and Davies, A.
(SUSSP publications, 1990);
C.-P. Ma and E. Bertschinger, ``Cosmological Perturbation Theory
in the Synchronous vs. Conformal Newtonian Gauge'',
MIT preprint (1994).
\bibitem{bardeen}
J. M. Bardeen, {\it Phys. Rev. D}{\bf 22} (1980) 1882.
\bibitem{kosa}
V.F. Mukhanov, H. A. Feldman and R. H. Brandenberger,
{\it Phys. Rep.} {\bf 215} (1992) 203;
H. Kodama and M. Sasaki, {\it Prog. Theor. Phys.} {\bf 78} (1984) 1;
N. Sugiyama, N. Gouda and M.
Sasaki, Astrophys. J. {\bf 365} (1990) 432 
and references cited there;
R. K. Schaefer, {\it Int. J. Mod. Phys.} {\bf A6} (1991) 2075;
W. Hu and N. Sugiyama, ``Anisotropies in the cosmic microwave
background'', Berkeley preprint (1994).
\bibitem{hawk}
S. W. Hawking, {\it Astrophys. J.} {\bf 145} (1966) 544;
D. W. Olson, {\it Phys. Rev. D}{\bf 14} (1976) 327.
\bibitem{ellis}
G. F. R. Ellis and M. Bruni, {\it Phys. Rev. D}{\bf  40} (1989) 1804;
M. Bruni, P. K. S. Dunsby and G. F. R. Ellis, 
{\it Astrophys. J.} {\bf 395} (1992) 34;
P. K. S. Dunsby, M. Bruni and G. F. R. Ellis, 
{\it Astrophys. J.} {\bf 395} (1992) 54.
\bibitem{lyth85} D. H. Lyth, {\it Phys. Rev. D} {\bf 31} (1985) 1792 .
\bibitem{brly} M. Bruni and D. H. Lyth, {\it Phys. Letts.}
{B 323} (1994) 118.
\bibitem{lymu} D. H. Lyth and M. Mukherjee, {\it Phys. Rev. D} {\bf 38} 
(1988) 485.
\bibitem{lyst}
D. H. Lyth and  E. D. Stewart, {\it Astrophys. J.} {\bf 361} (1990) 343.
\bibitem{lystomega} D. H. Lyth and E. D. Stewart, Phys Lett.
{\bf B252}, 336 (1993).
\bibitem{latest} E. J. Copeland, A. R. Liddle, D. H. Lyth, E. D. 
Stewart and D. Wands, {\it Phys. Rev. D} {\bf 49} (1994) 6410;
E. J. Copeland {\it et al}, this volume;
E. D. Stewart, ``Inflation, Supergravity and Superstrings'',
 Kyoto preprint (1994);
E. D. Stewart, ``Mutated Hybrid Inflation'', Kyoto preprint (1994).
\bibitem{sylvia}
S. Mollerach, S. Matarrese and F. Lucchin, ``Blue Perturbation
	Spectra from Inflation'', CERN preprint (1993).
\bibitem{harrison} R. Harrison, {\it Phys. Rev. D} {\bf 1} (1970) 2726.
Ya. B. Zel'dovich, {\it Astron. Astrophys.} {\bf 5} (1970) 84.
\bibitem{adpred}
A. A. Starobinsky, {\it Phys. Lett.} {\bf B117} (1982) 175;
S. W. Hawking, {\it Phys. Lett.} {\bf B115} (1982) 339;
A. H. Guth and S.-Y. Pi, {\it Phys. Rev. Lett.} {\bf 49} (1982) 1110;
J. M. Bardeen, P. S. Steinhardt and M. S. Turner,
{\it Phys. Rev. D} {\bf 28} (1983) 679.
\bibitem{mukhanov} V. F. Mukhanov, {\it JETP Lett.} {\bf 41} (1985) 493 .
\bibitem{sasaki} M. Sasaki, {\it Prog. Theor. Phys.} {\bf 76} (1986) 1036.
\bibitem{stly} Stewart, E. D., Lyth, D. H., {\it Phys. Lett.} {\bf B302}
(1993) 171.
\bibitem{nongauss} 
T. J. Allen, B. Grinstein and M. B. Wise, {\it Phys. Lett.} {\bf B197}
(1987) 66;
J. D. Barrow and P. Coles, {\it Mon. Not. R. ast. Soc.} {\bf 244 }(1990)
188;
I. Yi and E. T. Vishniac, {Phys. Rev. D} {\bf 48} (1993) 950;
T. Falk, R. Rangarajan and M. Srednicki, Astrophys. J. {\bf 403 }(1994) L1;
A. Gangui, F. Lucchin, S. Matarrese and S. Mollerach, {\it
Astrophys. J.} (1994) to appear;
A. Gangui, preprint (1994);
K. Yamamoto and M. Sasaki, preprint (1994).
\bibitem{starob} A. A.  Starobinsky, {\it Sov. Astron. Lett.} {\bf 11} 
(1985) 133.
\bibitem{sampvar} D. Scott, M. Srednicki and M. White, {\it Astrophys. J. 
Lett.} {\bf 421} (1994) L5.
\bibitem{cretal} R. Crittenden, J. R. Bond, R. L. Davis, G. Efstathiou,
P. J. Steinhardt and M. S. Turner, {\it Phys. Rev. Lett.} {\bf 71} (1993) 
324.
\bibitem{peebles82} P. J. E. Peebles, {\it Astrophys. J. Lett.} {\bf 263}
(1982) L1.
\bibitem{peadodd}
J. A. Peacock and S. J. Dodds, {\it Mon. Not. R. astr. Soc.} {\bf 267},
(1994) 1020.
\bibitem{dekelrev}
A. Dekel, ``Dynamics of Large-Scale Motions in the Universe'',
to appear in {\it Ann. Rev. Astron. Astroph.}, October (1994).
\bibitem{dekelet}
A. Dekel {\it et al}, {\it Astrophys. J.} {\bf 412} (1993) 1.
\bibitem{lacey}
C. Lacey and S. Cole, ``Merger Rates in Heirarchical Models of
        Galaxy Formation. II. Comparison with N-Body Simulations'', 
Oxford preprint (1994).
\bibitem{mabert2}
C.-P. Ma and E. Bertschinger, ``Do Galactic Systems Form Too
Late in Cold$+$Hot Dark Matter Models?'', preprint (1994).
\bibitem{whitecluster} S. D. M. White, G. Efstathiou and C. S. Frenk, 
{\it	Mon. Not. Roy. astr. Soc.} {\bf 262} (1993) 1023.
\bibitem{carlberg93} R. G. Carlberg {\em et al}, `Mapping
moderate redshift clusters', preprint (1993).
\bibitem{haehnelt}
M. G.  Haehnelt, {\it Mon. Not. Roy. astr. Soc.} {\bf 265} (1993) 727.
\bibitem{wolfe}
A. M. Wolfe, {\it Annals of the New York Academy of Sciences}
        {\bf 688} (1993) 281.
\bibitem{zurek} W. H. Zurek {\em et al}, in {\sl Proc. of the
9th IAP Meeting}, eds. F. R. Bouchet and M. Lachieze-Rey (Editions
Frontieres, gif-sur-Yvette, 1993), p. 465;
J. G. Bartlett and A. Blanchard, {\em ibid}, p. 281.
\bibitem{mojing} H. J. Mo, Y. P. Jing and G. B"{}orner, {\it Mon. Not. 
Roy. astr. Soc.} {\bf 264} (1993) 825.
\bibitem{dalton} G. B. Dalton {\em et al}, ``The two-point correlation 
function of rich clusters of galaxies'', Oxford preprint (1994).
\bibitem{natural1}K. Freese, J. A. Frieman and A. V. Olinto, 
{\it Phys. Rev. Lett.} {\bf 65} (1990) 3233. 
\bibitem{bobnext}
A. R. Liddle, D. H. Lyth, R. K. Schaefer, Q. Shafi and P. Viana,
in preparation (1994).
\bibitem{schneider} D. P. Schneider, M. Schmidt and J. E. Gunn, {\it Astron. 
J.} {\bf 98} (1989) 1507.
\bibitem{tegsilk}  M. Tegmark, J. Silk and A. Blanchard, {\it Astrophys. J.} 
{\bf 420} (1994) 484.
\end{thebibliography}
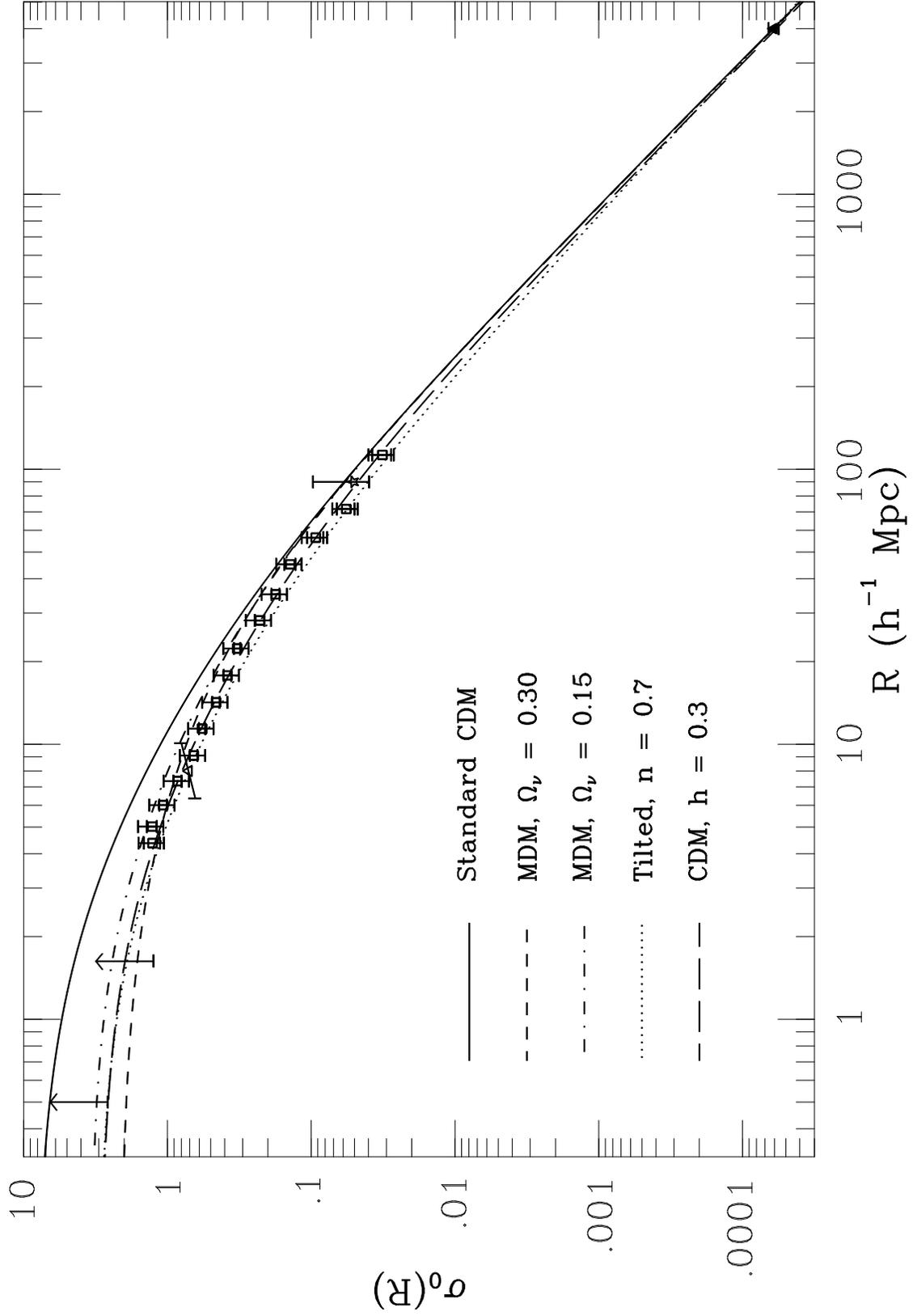
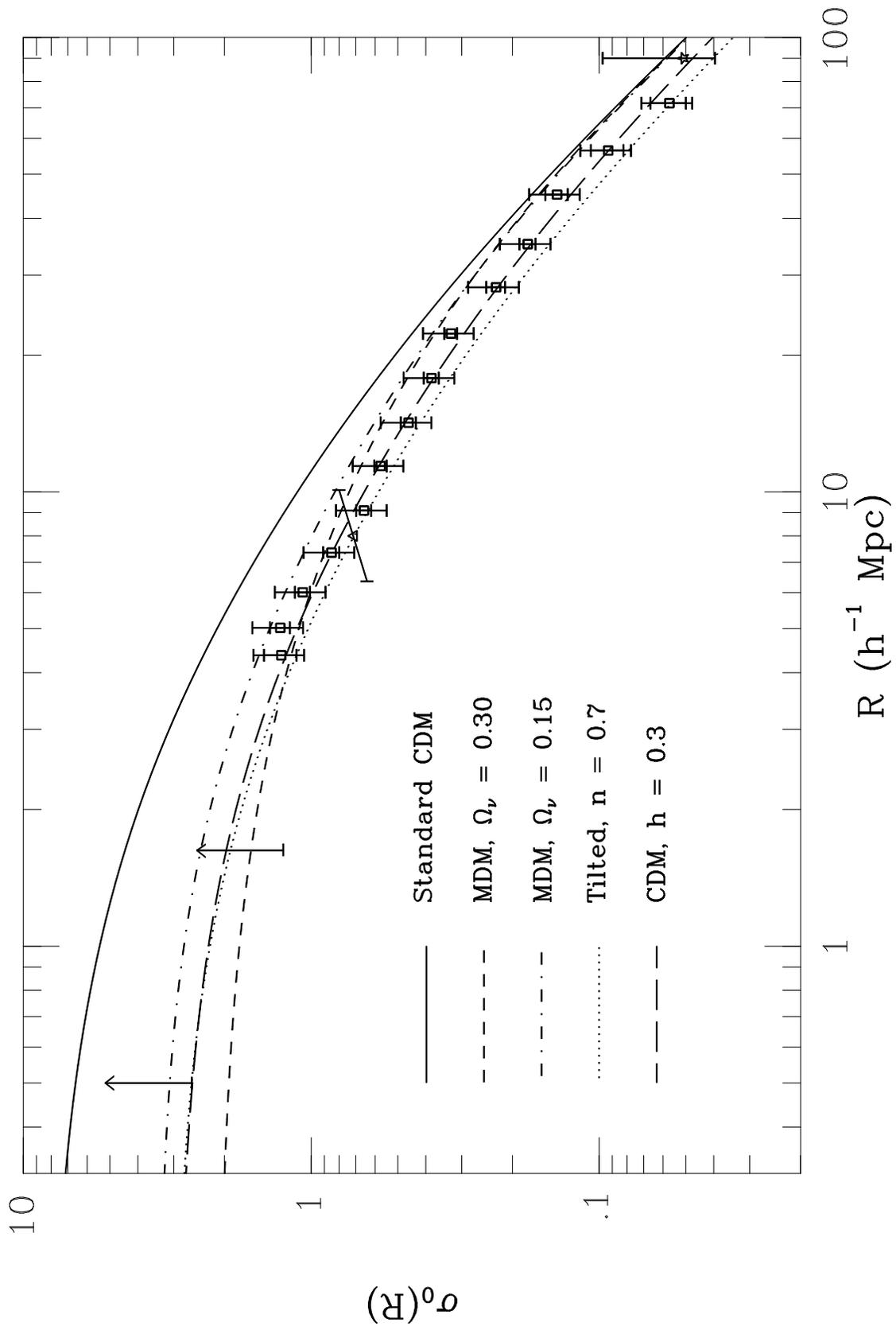
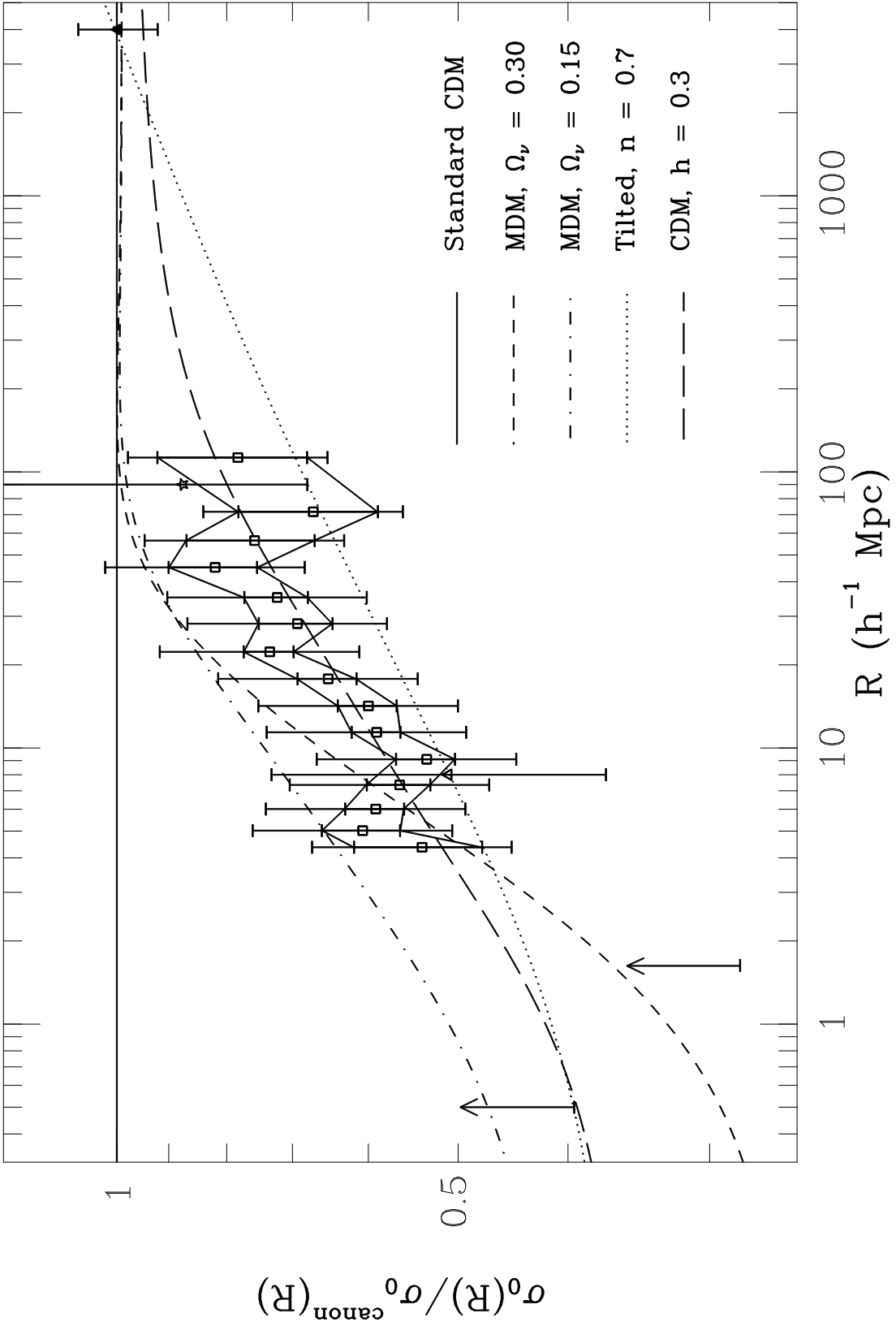
\end{document}